\documentclass[prd,nofootinbib,preprint]{revtex4}
\usepackage{graphicx}
\usepackage{epstopdf}
\usepackage{amsmath}
\usepackage{amsfonts}
\usepackage{amssymb}
\usepackage{url}
\usepackage{hyperref}
\usepackage{subfigure}

\newcommand{\nn}{\nonumber}
\newcommand{\bmt}{\begin{pmatrix}}
\newcommand{\emt}{\end{pmatrix}}
\usepackage[toc,page]{appendix}
\usepackage{comment}
\newcommand{\be}{\begin{equation}}
\newcommand{\ee}{\end{equation}}
\newcommand{\bea}{\begin{eqnarray}}
\newcommand{\eea}{\end{eqnarray}}

\begin{document}
\title{Model independent analysis of $ B^* \to P \ell \bar{\nu}_\ell$ decay processes}

\author{Atasi Ray$^a$}
\email{atasiray92@gmail.com}

\author{Suchismita Sahoo$^b$}
\email{suchismita8792@gmail.com}

\author{Rukmani Mohanta$^a$}
\email{rmsp@uohyd.ac.in}

\affiliation{$^a$School of Physics,  University of Hyderabad, Hyderabad-500046,  India\\
$^b$Theoretical Physics Division, Physical Research Laboratory, Ahmedabad-380009, India}

\begin{abstract}
Very compelling deviations  in the recently observed lepton nonuniversality  observables $\big (R_{D^{(*)}}, R_{K^{(*)}}, R_{J/\psi} \big )$ of  semileptonic $B$ meson decays  from their  Standard Model predictions   hint towards  the presence of some kind of new physics  beyond it. In this regard, we investigate the effect of new physics in the semileptonic  $\bar B_{d(s)}^* \to P \ell \bar{\nu}_\ell$ decay processes, where $P=D,\pi (D_s,K$), in a model independent way. We consider the presence of  additional vector and  scalar type interactions and   constrain the corresponding  new couplings  by fitting   ${\rm Br(B_{u}^+ \to \tau^+ \nu_\tau)}$, ${\rm Br(B \to \pi \tau \bar \nu_\tau)}$,  ${\rm Br(B_{c}^+ \to \tau^+ \nu_\tau)}$,  $R_\pi^l$,  $R_{D^{(*)}}$ and  $R_{J/\psi}$ data. Using the constrained new parameters, we estimate the branching ratios, forward-backward asymmetry, lepton-spin asymmetry and lepton non-universality observables of  $\bar B_{d,s}^{*} \to P \tau \bar \nu_\tau$ processes. We find that the branching ratios of these decay modes are sizeable and deviate significantly (for vector-type couplings)  from their corresponding standard model values, which are expected to be within the reach of Run III of Large Hadron Collider experiment.

\end{abstract}
\maketitle
\section{Introduction}

In the last few years, several intriguing hints of new physics (NP) have been observed  in the form of lepton flavour universality  violating (LFUV) observables  in semileptonic $B$ decays. In particular, the observables $R_{D^{(*)}}= {\rm Br}(B \to D^{(*)} \tau \bar \nu_\tau)/{\rm Br}(B \to D^{(*)} l \bar \nu_l) $, with $l=e,\mu$  in the  charged-current transition $b \to c \ell \bar \nu_\ell$, measured by BaBar \cite{Lees:2012xj,Lees:2013uzd}  Belle  \cite{Huschle:2015rga, Sato:2016svk, Hirose:2016wfn,Abdesselam:2019dgh} and LHCb \cite{Aaij:2015yra, Aaij:2017uff, Aaij:2017deq} Collaborations, with the following  avarage values as determined by Heavy Flavour Averaging Group (HFLAV)   \cite{Hfag:2019}
\bea
R_D=0.340 \pm 0.027 \pm 0.013\;,~~~~~R_{D^*}=0.295 \pm 0.011 \pm 0.008 \;,
\eea
with $R_D-R_{D^*}$ correlation of $-0.38$, 
indicate $\sim 3.08 \sigma $ discrepancy  with their  corresponding Standard Model (SM) predictions
\bea
R_D^{\rm SM}=0.299 \pm 0.003\;,~~~~R_{D^*}^{\rm SM}=0.258 \pm 0.005\;.
\eea 
The recently measured $R_{J/\psi}={\rm Br}(B_c \to J/\psi \tau \bar \nu_\tau)/{\rm Br}(B_c \to J/\psi l \bar \nu_l) =0.71\pm0.17\pm0.184$ parameter by LHCb Collaboration  \cite{Aaij:2017tyk} is in the same line and has nearly $2\sigma$ deviation from its SM value $R_{J/\psi}=0.289 \pm 0.01$ \cite{Wen-Fei:2013uea, Ivanov:2005fd}.  Similarly, in the  semileptonic $B \to K^{(*)} \ell \ell$ decay processes,  mediated by the neutral current  transition $b \to s \ell \ell$,  $2.6\sigma $  and $(2.2-2.4)\sigma $ deviations have been  observed in the measured values of $R_{K}={\rm Br}(B^+ \to K^+ \mu^+ \mu^-)/{\rm Br}(B^+ \to K^+ e^+ e^-)$ \cite{Aaij:2014ora} and $R_{K^*}={\rm Br}({\bar B}^0 \to {\bar K}^* \mu^+ \mu^-)/{\rm Br}({\bar B}^0 \to {\bar K}^* e^+ e^-)$  \cite{Aaij:2017vbb} with values
\bea
&&R_K|_{q^2\in [1, 6]~{\rm GeV}^2} = 0.745^{+0.090}_{-0.074} \pm 0.036\;,\nn\\
&& R_{K^*}|_{q^2\in [0.045,1.1]~{\rm GeV}^2} =0.66^{+0.11}_{-0.07} \pm 0.03\;,~~~
  R_{K^*}|_{q^2\in [1.1, 6]~{\rm GeV}^2} =0.69^{+0.11}_{-0.07} \pm 0.05\;, 
\eea
 from their corresponding SM predictions \cite{Bobeth:2007dw, Capdevila:2017bsm}
 \bea
&& R_K^{\rm SM}|_{q^2\in [1, 6]~{\rm GeV}^2} =1.003\pm 0.0001\;,\nn\\
&& R_{K^*}{\rm SM}|_{q^2\in [0.045,1.1]~{\rm GeV}^2}=0.92\pm 0.02\;,~~~~
  R_{K^*}^{\rm SM}|_{q^2\in [1.1, 6]~{\rm GeV}^2} = 1.00 \pm 0.01.
  \eea
Recently, the LHCb experiment has announced its updated measurements on $R_K$ \cite{Aaij:2019wad} and the Belle Collaboration has announced new $R_{K^*}$ \cite{Prim:2019hyn,Prim:2019wem} results.  After combining the Run 1 and Run 2 data, though the updated experimental value of $R_K=0.846^{+0.060}_{-0.054}{\rm (stat)}^{+0.016}_{-0.014}{\rm (syst)}$ \cite{Aaij:2019wad} is closer to the SM prediction, the discrepancy still persists at the level of  $\sim 2.5\sigma$, due to the reduced errors. The errors in the  new measurements on $R_{K^*}=0.52^{+0.36}_{-0.26}\pm 0.005~(0.96^{+0.45}_{-0.29}\pm 0.11)$ observable in the $q^2\in [0.045,1.1]~{\rm GeV}^2~(q^2\in [1.1,6]~{\rm GeV}^2)$ bin,  reported by the Belle Collaboration  \cite{Prim:2019hyn,Prim:2019wem} are quite a bit larger than the errors in the previous LHCb masurement.    Additionly, a small discrepancy   has also been  reported in  the $b \to u \ell \bar \nu$ mediated process  defined as $R_\pi^{l }=\frac{\tau_{B^0}}{\tau_{B^-}}({\rm{Br}(B^-\to \tau^- \bar{\nu}_\tau)}/{\rm{Br}}(B^0\to \pi^ +l^- \bar{\nu}_l))$ \cite{Tanabashi:2018oca}. As all these  observables are ratios of branching fractions, the theoretical uncertainties due to the CKM matrix elements and hadronic form factors  cancel out to a large extent, resulting the prediction with high accuracy. Therefore,  the lepton flavor universality violating  tests are considered to be the most powerful tools to probe  new physics  beyond the standard model.    Tremendous effort has been made in the last few years to understand the nature of NP,  which might be responsible for such deviations.

Being motivated by these observed anomalies in various   $B$ meson decays, in this work we would like to investigate the impact of new physics on the differential decay rate and various other observables like forward-backward asymmetry, lepton-spin asymmetry and lepton nonuniversality (LNU) observable of weakly decaying vector $B_{d,(s)}^*$ meson to a pseudoscalar $P(=D (D_s), \pi (K))$ meson  mediated through the quark level transitions $b \to (c,u) \ell \bar \nu_\ell$.  Although such hadrons decay primarily through the electromagnetic process $B_{d,s}^* \to B_{d,s} \gamma$, and their weak decay channels are  expected to be quite suppressed, the situation has improved considerably with the advent of the high  luminosity Belle II experiment.  For instance,  as discussed in Ref. \cite{Chang:2016cdi},   using the production cross section of $\Upsilon(5S)$ in $e^-e^+$ collision as  $\sigma(e^+ e^- \to \Upsilon(5S))=0.301 $nb  and  ${\rm Br}(\Upsilon(5S) \to  B^* \bar B^*)=(38.1\pm 3.4)\%$ \cite{Tanabashi:2018oca},  about $ 4 \times 10^9$ $B^*$ meson pairs ($ B_{u,d}^* + \bar B_{u,d}^*$)  are expected to be produced per year. This in turn implies that the rare $B^* $ decay modes with branching fraction $>{\cal O}(10^{-9})$ are likely to be observed at Belle II. Hence,  Belle II experiment would be quite instrumental in search for the rare decay modes of the excited $B$ mesons.   In addition the LHC experiment will also play a pivotal role in the search for $B^*$ decay channels, as the production cross section of $\Upsilon(5S)$ is much larger in $p \bar p$ collision compared to $e^+e^-$ collision.  On the other hand, the study of  $B^*$ meson decays has also received considerable attention in recent times. In the literature  \cite{Grinstein:2015aua, Sahoo:2016edx, Kumar:2017xgl, Kumbhakar:2018uty}, the leptonic decay modes of $B_{s,d}^*$ mesons are investigated in SM and in the context of various new physics models. The analysis of semileptonic weak decays $B^* \to P \ell \nu$ both in the SM and in the   presence of NP are  discussed in the Refs. \cite{Chang:2016cdi,Chang:2018sud,Zhang:2019hth}.

The layout of the paper is as follows. In section II, we illustrate the theoretical framework required to analyse the decay processes $B^* \to P \ell \nu$ in the effective theory formalism. The expressions for the differential decay rate and other observables like forward-backward asymmetry, lepton nonuniversality   $(R_P^*)$ and the lepton-spin asymmetry  are presented in this section.
The constraints on the new couplings using $\chi^2$ fit from $R_{D^{(*)}}$, $R_{J/\psi}$, $R_\pi^l$, Br($B_{u,c} \to \tau \nu$), Br($B \to \pi \tau \bar \nu$) observables are obtained in section III. Our results are discussed  in  section IV followed by the  summary of our work in section V.

\section{Theoretical Framework}
The most general effective Lagrangian for $B^*\to P \ell \bar{\nu}_\ell$ processes mediated by $b\to q \ell^- \bar{\nu}_\ell$ ($q=u,c$), in the effective field theory approach can be expressed as \cite{Tanaka:2012nw},
\begin{eqnarray}
\mathcal{L}_{eff} &=&-2\sqrt{2}G_F  V_{q b}\Big[(1+V_L)~\bar{q}_L\gamma^\mu b_L~ \bar{\ell}_L \gamma_\mu \nu_L +V_R~\bar{q}_R\gamma^\mu b_R ~\bar{\ell}_L \gamma_\mu \nu_L +S_L~\bar{q}_R b_L~\bar{\ell}_R\nu_L\nonumber\\ && +S_R~\bar{q}_L b_R~\bar{\ell}_R\nu_L 
+T_L~\bar{q}_R \sigma^{\mu \nu} b_L~\bar{\ell}_R \sigma_{\mu \nu}\nu_L +{\rm h.c.}\Big],\label{Lag}
\end{eqnarray}
where $P$ is any pseudoscalr meson, $G_F$ is the Fermi constant, $V_{q b}$ is the CKM matrix element, $V_{L,R},~ S_{L,R}, T_L$ are the new vector, scalar, and tensor type  new physics couplings, which are zero in the standard model. All these new physics  couplings are considered to be complex. Furthermore, we consider the neutrinos as left handed. We assume the NP effect is mainly through the third generation leptons and do not consider the effect of  tensor operators in our analysis for simplicity.   Here $(q,\ell)_{L,R}=P_{L,R}(q,\ell)$, where $P_{L,R}=(1\mp \gamma_5)/2$ are the chiral projection operators. 

We consider the kinematics of the decay process $B^* \to P \ell \bar \nu_\ell$ using helicity amplitudes. In this formalism, the decay process $B^* \to P \ell \bar \nu_\ell$  is considered to proceed through  $\bar B^* \to P W^{*-}$, where the off-shell $W^{*-}$ decays to $\ell^- \bar \nu_\ell$. One can write the amplitude from Eq. (\ref{Lag}) as
\bea
{\cal M}(B^* \to P \ell \bar \nu_\ell)= \frac{G_F}{\sqrt 2} V_{qb}\sum_k  C_k(\mu) \langle P | \bar q \Gamma^k b |B^* \rangle~ \bar u_\ell \Gamma_k v_\nu\;,
\eea
where $C_k(\mu)$ represents the Wilson coefficient with values  \[ 
  C_k(\mu) = \begin{cases}
  1  & {\rm for~SM}\,, \\ 
  V_{L,R},S_{L,R}  & {\rm for~NP~beyond~SM}\,,
   \end{cases}
\] 
 $\Gamma^k$ denotes the product of gamma matrices, which gives rise to different Lorentz structure of hadronic  and leptonic  currents  of Eq. (\ref{Lag}) i.e.,
$\Gamma^k = \gamma^\mu(1\pm \gamma_5)$, and~$(1\pm \gamma_5) $. Hence, the square of the matrix element can be expressed as the product of leptonic ($L_{\mu \nu}$) and hadronic $(H^{\mu \nu})$ tensors (related to the corresponding helicity amplitudes)
\bea
\big |{\cal M}(B^* \to P \ell \bar \nu_\ell)\big  |^2= \frac{G_F^2}{ 2}| V_{qb}|^2 \sum_{i,j}C_{ij}(\mu)   \Big (L_{\mu \nu}^{ij}H^{\mu \nu,ij}\Big )\;,\label{mat-amp}
\eea 
where the superscripts $i,j$ represent the combination of four operators $(V\mp A), (S\mp P)$ in the effective Lagrangian (\ref{Lag}), $C_{ij}(\mu)$ denotes the product of Wilson coefficients $C_i$ and $C_j$. We omit these superscripts in the following discussion for convenience. It should be noted that, the  polarization vector of the  off-shell particle $W^*$ ($\bar \epsilon^{\mu}(m))$, satisfies the
following orthonormality and completeness relations:
\bea
&&\bar \epsilon^{* \mu} (m) \bar \epsilon_\mu(m')=g_{m m'}\;,\nn\\
 && \sum_{m m'}\bar \epsilon^{* \mu} (m) \bar \epsilon^\nu(m') g_{m m'}=g^{\mu \nu}\;,\label{complete}
 \eea
 where $g_{m m'}={\rm diag}(+,-,-,-)$ and 
  $m,m'=\pm,0,t$ represent the transverse, longitudinal and time-like polarization components. Now inserting the completeness relation from Eq. (\ref{complete}) into (\ref{mat-amp}), the product of $L_{\mu \nu}$ and  $H^{\mu \nu}$ can be expressed as
  \bea
  L_{\mu \nu} H^{\mu \nu} = \sum_{m,m',n,n'} L(m,n) H(m',n') g_{mm'} g_{nn'}\;,
  \eea
  where $L(m,n)=L^{\mu \nu} \bar \epsilon_\mu (m)\bar \epsilon_\nu^*(n)$ and 
  $H(m,n)=H^{\mu \nu} \bar \epsilon_\mu^* (m)\bar \epsilon_\nu(n)$ are the Lorentz invariant parameters, and hence their values are independent of any specific reference frame. So for calculational convenience, we will evaluate $H(m,n)$ in the $B^*$ rest frame and $L(m,n)$ in $\ell-\bar \nu_\ell$ center of mass frame as discussed in \cite{Chang:2018sud, Chang:2016cdi}.
\subsection{Hadronic helicity amplitudes} 
In the rest frame of $B^*$ meson, we consider the pseudoscalar meson $P$ to be moving along the positive $z$-direction. The polarization vector of the virtual $W^*$ boson are chosen to be  
 \begin{equation}
\bar{\epsilon}^\mu(t)=\frac{1}{q^2}(q_0,0,0,-|\vec{p}|), ~~ \bar{\epsilon}^\mu(0)=\frac{1}{q^2}(|\vec{p}|,0,0,-q_0), ~~ \bar{\epsilon}^\mu(\pm)=\frac{1}{\sqrt{2}}(0,\pm 1,-i,0),
\end{equation}
where $q_0=(m_{B^*}^2-m_P^2+q^2)/{2m_{B^*}}$,  $|\vec{p}|=\lambda^{1/2}(m^2_{B^*},m_P^2,q^2)/{2m_B^*}$,  $q^2=(p_{B^*}-p_P)^2$,  is the momentum transferred square and $\lambda(a,b,c)=a^2+b^2+c^2-2(ab+bc+ca)$.
The polarization vector of the on-shell $B^*$ meson $\varepsilon^\mu(m=0,\pm),$ takes the form
\begin{equation}
\varepsilon^\mu (0)=(0,0,0,1),~~ \varepsilon^\mu (\pm)=\frac{1}{\sqrt{2}}(0,\mp 1,-i,0)\;.
\end{equation}
In order to calculate the hadronic helicity amplitudes, we use the following matrix elements of $B^* \to P$ transition 
\begin{eqnarray}
\langle P(p_P)|\bar{q}\gamma_\mu b |\bar{B^*}(\varepsilon,p_{B^*})\rangle &=&-\frac{2iV(q^2)}{m_{B^*}+m_P}\epsilon_{\mu\nu\rho\sigma}\varepsilon^\nu p_P^\rho p_{B^*}^\sigma , \nonumber \\
\langle P(p_P)|\bar{q}\gamma_\mu \gamma_5 b|\bar{B^*}(\varepsilon,p_{B^*})\rangle&=&2m_{B^*}A_0(q^2)\frac{\varepsilon \cdot q}{q^2}q_\mu +(m_P+m_{B^*})A_1(q^2)(\varepsilon_\mu-\frac{\varepsilon\cdot q}{q^2}q_\mu)\nonumber\\
&+&A_2(q^2)\frac{\varepsilon \cdot q}{m_P+m_{B^*}}\big[(p_{B^*}+p_P)_\mu -\frac{m_{B^*}^2-m_P^2}{q^2}q_\mu \big]\;,\label{form}
\end{eqnarray}
where $V(q^2),~ A_{0,1,2}(q^2)$ are the various form factors.
The matrix elements for the scalar and pseudoscalar currents can be obtained by using the equation of motion
\bea
i \partial_\mu (\bar q \gamma^\mu b)=(m_b -m_q)\bar qb\;,~~~~~i \partial_\mu (\bar q \gamma^\mu\gamma_5 b)=-(m_b+m_q)\bar q\gamma_5 b\;,
\eea
as
\begin{eqnarray}
\langle P(p_P)|\bar{q}b|\bar{B^*}(\varepsilon,p_{B^*})\rangle &=& 0\;,\nonumber\\
\langle P(p_P)|\bar{q} \gamma_5 b|\bar{B^*}(\varepsilon,p_{B^*})\rangle &=& 
-(\varepsilon .q)\frac{2m_{B^*}}{m_b+m_{q}}A_0(q^2),\label{form-1}
\end{eqnarray}
where the $m_{b,q}$ represent the current quark masses evaluated at the $b-$quark mass scale. 
The helicity amplitudes are defined as
\begin{eqnarray}
H_{\lambda_{B^*}\lambda_{W^*}}^{V_L}(q^2)&=&\bar{\epsilon}^{*\mu}(\lambda_{W^*})\langle P(p_P)|\bar{q}\gamma_\mu (1-\gamma_5)b|\bar{B}^*(\varepsilon(\lambda_{B^*}),p_{B^*})\rangle ,\nn\\
H_{\lambda_{B^*}\lambda_{W^*}}^{V_R}(q^2)&=&\bar{\epsilon}^{*\mu}(\lambda_{W^*})\langle P(p_P)|\bar{q}\gamma_\mu (1+\gamma_5)b|\bar{B}^*(\varepsilon(\lambda_{B^*}),p_{B^*})\rangle ,\nn\\
H_{\lambda_{B^*}\lambda_{W^*}}^{S_L}(q^2)&=&\langle P(p_P)|\bar{q} (1-\gamma_5)b|\bar{B}^*(\varepsilon(\lambda_{B^*}),p_{B^*})\rangle ,\nn\\
H_{\lambda_{B^*}\lambda_{W^*}}^{S_R}(q^2)&=&\langle P(p_P)|\bar{q}\ (1+\gamma_5)b|\bar{B}^*(\varepsilon(\lambda_{B^*}),p_{B^*})\rangle ,\label{helicity}
\end{eqnarray}
where for convenience, we use the notations  $\lambda_{B^*}=0,\pm$ and $\lambda_{W^*}=0,\pm,t$  to represent  the helicity states of the $B^*$ and $W^*$ boson.  Thus, with Eqs. (\ref{form}), (\ref{form-1}) and (\ref{helicity}), one obtains the following non-vanishing helicity amplitudes
  \begin{eqnarray}
H_{0t}(q^2)&=&H_{0t}^{V_L}(q^2)=-H_{0t}^{V_R}(q^2)=\frac{2m_{B^*}|\vec p|}{\sqrt{q^2}}A_0(q^2)\;,\nn\\
H_{00}(q^2)&=&H_{00}^{V_L}(q^2)=-H_{00}^{V_R}(q^2)\nonumber\\
&=&\frac{1}{2m_{B^*}\sqrt{q^2}}\Big[(m_{B^*}+m_P)(m_{B^*}^2-m_P^2+q^2)A_1(q^2)+\frac{4m_{B^*}^2|\vec p|^2}{m_{B^*}+m_P}A_2(q^2)\Big]\;,\nn\\
H_{\pm \mp}(q^2)&=&H_{\pm \mp}^{V_L}(q^2)=-H_{\mp \pm}^{V_R}(q^2)=-(m_{B^*}+m_P)A_1(q^2)\mp \frac{2m_{B^*}|\vec p|}{m_{B^*}+m_P}V(q^2)\;,\nn\\
H^\prime_{0t}&=&H_{0t}^{S_L}(q^2)=-H_{0t}^{S_R}(q^2)=-\frac{2m_{B^*}|\vec p|}{m_b+m_{q}}A_0(q^2)\;.\label{hadronic}
\end{eqnarray}
  
\subsection{Leptonic helicty amplitudes}
  The leptonic helicity amplitudes are defined as
  \bea
  h_{\lambda_\ell, \lambda_{\bar \nu_\ell}}^i= \frac{1}{2}\bar \epsilon_\mu(\lambda_{W^*})
 ~ \bar u_\ell(\lambda_\ell) ~\Gamma^i ~v_{\bar \nu_\ell}(\lambda_{\bar \nu_{\ell}})\;,\label{lepton}
  \eea
  where $\lambda_{W^*}= \lambda_\ell-\lambda_{\bar \nu_\ell}$, $i=V_{L,R},S_{L,R}$, and $\Gamma^{V_{L,R}}=\gamma^\mu(1 \mp \gamma_5)$, $\Gamma^{S_{L,R}}=(1\mp \gamma_5)$. In the center of mass frame of $\ell-\bar \nu_\ell$, the four momenta of $\ell$ and $\bar \nu_\ell$ pair are expressed as
  \bea
  p_\ell^\mu=\big (E_\ell, |\vec p_\ell | \sin \theta, 0, |\vec p_\ell | \cos \theta\big ),~~~~ p_{\nu_\ell}^\mu=\big (|\vec p_\ell |, -|\vec p_\ell | \sin \theta, 0, -|\vec p_\ell | \cos \theta\big )\;,
  \eea
  where $E_\ell=(q^2+m_\ell^2)/2 \sqrt{q^2}$,  $|\vec p_\ell |=(q^2-m_\ell^2)/2 \sqrt{q^2}$ and $\theta$ is the angle between the three momenta of of $P$ and $\ell$. The polarization vector of the virtual $W^*$ boson in this frame is 
  \bea
  \bar \epsilon^\mu(t)=(1,0,0,0),~~~~\bar \epsilon^\mu(0)=(0,0,0,1),~~~~~\bar \epsilon^\mu(\pm)=\frac{1}{\sqrt{2}}(0, \mp1, -i,0)\;.\label{lepton-1}
  \eea 
  Thus, with Eqs. (\ref{lepton}) and (\ref{lepton-1}), one obtains the following non-vanishing contributions
  \bea
 && |h_{-\frac{1}{2}, \frac{1}{2}}^{V_{L,R}}|^2= 8(q^2-m_\ell^2),~~~~~~~ 
   |h_{\frac{1}{2}, \frac{1}{2}}^{V_{L,R}}|^2= 8\frac{m_\ell^2}{2 q^2}(q^2-m_l^2),\nn\\
   &&  |h_{\frac{1}{2}, \frac{1}{2}}^{S_{L,R}}|^2= 4(q^2-m_\ell^2),~~~~~~~ 
   |h_{\frac{1}{2}, \frac{1}{2}}^{V_{L,R}}| \times  |h_{\frac{1}{2}, \frac{1}{2}}^{S_{L,R}}|= 8\frac{m_\ell}{2\sqrt{ q^2}}(q^2-m_\ell^2)\;.\label{leptonic}
  \eea
  \subsection{Decay distribution and other observables}  
The double differential decay rate of $B^*\to P \ell\bar{\nu}_\ell$ decay process can be expressed  as 
\begin{eqnarray}
\frac{d^2 \Gamma}{dq^2 d\cos\theta}=\frac{G_F^2}{192 \pi^3} \frac{|\vec{p}|}{ m_{B^*}^2} |V_{q b}|^2\left (1-\frac{m_\ell^2}{q^2}\right )\big |\mathcal{M}(\bar{B}^*\to P \ell \bar{\nu}_\ell)\big |^2\;.
\end{eqnarray}  
  Now, with Eqs. (\ref{hadronic}) and (\ref{leptonic}), one can obtain $L_{\mu \nu} H^{\mu \nu}$ in terms  of Wigner $d^J$-functions as \cite{Chang:2018sud}
  \bea
  L_{\mu \nu}H^{\mu \nu}&=&\frac{1}{8}\sum_{\lambda_\ell, \lambda_\nu, \lambda_{W^*}, \lambda'_{W^*},J,J'} (-1)^{J+J'} h^i_{\lambda_\ell, \lambda_\nu}
  h^{j*}_{\lambda_\ell, \lambda_\nu} \delta_{\lambda_{B^*},-\lambda_{W^*}}
  \delta_{\lambda_{B^*},-\lambda'_{W^*}}\\
  && \times  d^J_{\lambda_{W^*}, \lambda_{\ell-{1/ 2}}} d^{J'}_{\lambda'_{W^*}, \lambda_{\ell-{1/ 2}}} H^i_{\lambda_{B^*} \lambda_{W^*}} H^{j*}_{\lambda_{B^*} \lambda'_{W^*}}\;,
  \eea
  where $J$ and $J'$ take the values 0 and 1 and the various helicity components run over their allowed values. Thus, one can obtain the 
the differential decay rate to particular leptonic helicity state $(\lambda=\pm {1 \over 2})$  as
\begin{eqnarray}
\frac{d^2\Gamma(\lambda_\ell =-\frac{1}{2})}{dq^2d\cos\theta}&=&\frac{G_F^2 }{768\pi^3} \frac{ |\vec p|}{m_{B^*}^2} |V_{q b}|^2~q^2\left(1-\frac{m_\ell^2}{q^2}\right )^2 \Big\{|1+V_L|^2\big[(1-\cos\theta )^2H_{-+}^2 +(1+\cos\theta )^2 H_{+-}^2\nonumber\\
&+&2 \sin^2\theta H_{00}^2\big]+|V_R|^2\big[(1-\cos\theta)^2H_{+-}^2+(1+\cos\theta)^2H_{-+}^2+2\sin^2\theta H_{00}^2\big]\nonumber\\
&-&4\mathcal{R}e \big [(1+V_L)V_R^*]\big[(1+\cos\theta)^2H_{+-}H_{-+}+\sin^2\theta H_{00}^2 \big ]\Big\}\;,\label{br-1}
\end{eqnarray}
\begin{eqnarray}
\frac{d^2\Gamma(\lambda_\ell =\frac{1}{2})}{dq^2d\cos \theta }&=&
\frac{G_F^2 }{768\pi^3} \frac{ |\vec p|}{m_{B^*}^2} |V_{q b}|^2~
\left(1-\frac{m_\ell^2}{q^2}\right)^2 m_\ell^2 \Big\{ \big(|1+V_L|^2+|V_R|^2\big)\big[\sin^2 \theta (H^2_{-+}+H^2_{+-})\nn\\
&+&2(H_{0t}-\cos \theta H_{00})^2\big]
-4\mathcal{R}e\big[(1+V_L)V_R^* \big]\big[\sin^2 \theta H_{-+}H_{+-}+(H_{0t}-\cos\theta H_{00})^2\big]\nonumber\\
&+&4\mathcal{R}e[(1+V_L-V_R)(S_L^*-S_R^*)]\frac{\sqrt{q^2}}{m_\ell}\big [H_{0t}^\prime(H_{0t}-\cos\theta H_{00})\big ]\nonumber\\
&+&2|S_L-S_R|^2\frac{q^2}{m_\ell^2}{H_{0t}^\prime}^2\Big\}.\label{br-2}
\end{eqnarray}
From Eqs. (\ref{br-1}) and (\ref{br-2}), one can obtain  the differential decay rate as
\begin{eqnarray}
\frac{d\Gamma}{dq^2}&=& \frac{G_F^2 }{288\pi^3} \frac{ |\vec p|}{m_{B^*}^2} |V_{q b}|^2 ~q^2 \left (1-\frac{m_\ell^2}{q^2}\right )^2\Big[(|1+V_L|^2+|V_R|^2)\nonumber\\
&\times &\big[\left (H_{-+}^2+H_{+-}^2+H_{00}^2\right )\left (1+\frac{m_\ell^2}{2q^2}\right )+\frac{3m_\ell^2}{2q^2}H_{0t}^2\big]\nonumber\\
&-&2\mathcal{R}e[(1+V_L)V_R^*]\Big [(2H_{-+}H_{+-}+H_{00}^2)\left (1+\frac{m_\ell^2}{2q^2}\right )+\frac{3m_\ell^2}{2q^2}H_{0t}^2\Big]\nonumber\\
&+&3\frac{m_\ell}{\sqrt{q^2}}\mathcal{R}e\Big[(1+V_L- V_R)(S_L^*-S_R^*)\Big]H^\prime_{0t}H_{0t}+\frac{3}{2}|S_L-S_R|^2H^\prime_{0t}\Big],
\end{eqnarray}
where the values of the helicity amplitudes are given in Eq. (\ref{hadronic}). 

Apart from the differential decay rate, the other NP sensitive observables, considered here  are
\begin{itemize}
\item Lepton nonuniversality observable:
\bea
R_P^*(q^2)=\frac{d\Gamma(B^*\to P\tau^-\bar{\nu}_\tau) /dq^2}{d\Gamma(B^*\to P l^-\bar{\nu}_l) /dq^2}
\;,
\eea
where $l$ denotes the light leptons $l=e,\mu$.
\item Forward-backward asymmetry:
\bea
A_{\rm FB}^P(q^2)=\frac{\int_{-1}^0 d\cos\theta(d^2\Gamma/dq^2d\cos\theta)-\int_{0}^1 d\cos\theta(d^2\Gamma/dq^2d\cos\theta)}
{\int_{-1}^0 d\cos\theta(d^2\Gamma/dq^2d\cos\theta)+\int_{0}^1 d\cos\theta(d^2\Gamma/dq^2d\cos\theta)}\;,
\eea
which can be expressed in terms of the helicity amplitudes as
\bea
A_{\rm FB}^P(q^2)=\frac{3}{4} \frac{X}{Y}\;,
\eea
where the parameters $X$ and $Y$ are given as
\bea
X&=&\left (|1+V_L|^2-|V_R|^2 \right )\left (H_{-+}^2 -H_{+-}^2  \right )+2\left (\frac{m_\ell^2}{q^2}\right )\big (|1+V_L|^2+|V_R|^2 \big )H_{0t}H_{00}\nn\\
&+&4 \mathcal{R}e[(1+V_L)V_R^*]\left( H_{+-}H_{-+}- \frac{m_\ell^2}{q^2} H_{0t}H_{00}\right)\nn\\
&+&2\mathcal{R}e[(1+V_L-V_R)(S_L^*-S_R^*)]\frac{m_\ell}{\sqrt{q^2}}H'_{0t}H_{00}\;,\nn\\
Y&=&\left (|1+V_L|^2+|V_R|^2\right )\big[\left (H_{-+}^2+H_{+-}^2+H_{00}^2\right )\left (1+\frac{m_\ell^2}{2q^2}\right )+\frac{3m_\ell^2}{2q^2}H_{0t}^2\big]\nonumber\\
&-&2\mathcal{R}e[(1+V_L)V_R^*]\Big[(2H_{-+}H_{+-}+H_{00}^2)\left (1+\frac{m_\ell^2}{2q^2}\right )+\frac{3m_\ell^2}{2q^2}H_{0t}^2\Big]\nonumber\\
&+&3\frac{m_\ell}{\sqrt{q^2}}\mathcal{R}e\Big[(1+V_L- V_R)(S_L^*-S_R^*)\Big]H^\prime_{0t}H_{0t}+\frac{3}{2}|S_L-S_R|^2H^\prime_{0t}\;.
\eea
 
\item Lepton-spin asymmetry:
\bea
A_\lambda^P(q^2)=\frac{d\Gamma(\lambda_\ell=-1/2)/dq^2-d\Gamma(\lambda_\ell=1/2)/dq^2}{d\Gamma(\lambda_\ell=-1/2)/dq^2+d\Gamma(\lambda_\ell=1/2)/dq^2}.
\eea
\end{itemize}

\subsection{Form factors and their $q^2$ dependence}
The main inputs required for the numerical analysis are the values of the form factors. As the first principle lattice calculation results of the form factors for $B_{d,s}^* \to D,D_s (\pi,K)$  transitions are not yet available, we use their values evaluated in the BSW model   \cite{Wirbel:1985ji,Bauer:1986bm}. Their values at zero-momentum transfer are listed below
\begin{eqnarray}
&&A_0^{\bar{B}^*\to D}(0)=0.71, ~~A_1^{\bar{B}^*\to D}(0)=0.75,~~
 A_2^{\bar{B}^*\to D}(0)=0.62, ~~V^{\bar{B^*}\to D}(0)=0.76,\nn\\
 &&A_0^{\bar{B}_s^*\to D_s}(0)=0.66, ~~A_1^{\bar{B}_s^*\to D_s}(0)=0.69,~~
 A_2^{\bar{B}_s^*\to D_s}(0)=0.59, ~~V^{\bar{B}_s^*\to D_s}(0)=0.72,\nn\\
 &&A_0^{\bar{B}^*\to \pi}(0)=0.34,~~ A_1^{\bar{B}^*\to \pi}(0)=0.38,
 ~~~A_2^{\bar{B}^*\to \pi}(0)=0.30,~~~ V^{\bar{B}^*\to \pi}(0)=0.35,\nn\\
  &&A_0^{\bar{B}_s^*\to K}(0)=0.28,~~ A_1^{\bar{B}_s^*\to K}(0)=0.29,
 ~~~A_2^{\bar{B}_s^*\to K}(0)=0.26,~~~ V^{\bar{B}_s^*\to K}(0)=0.30.
\end{eqnarray}
The $q^2$ dependence of the form factors can be written as,
\begin{eqnarray}
&&A_0(q^2)\simeq \frac{A_0(0)}{1-{q^2}/{m_{B_q}^2(0^-)}},~~~A_1(q^2)\simeq\frac{A_1(0)}{1-{q^2}/{m_{B_q}^2(1^+)}},\nonumber \\
&&A_2(q^2)\simeq\frac{A_2(0)}{1-{q^2}/{m_{B_q}^2(1^+)}},~~~V(q^2)\simeq \frac{V(0)}{1-{q^2}/{m_{B_q}^2(1^-)}}\;,
\end{eqnarray}
 where $m_{B_q}(0^\pm)$ and $ m_{B_q}(1^\pm)$  are the pole masses whose values are presented in Table \ref{Tab:polemass}. In our analysis, we consider $10\%$ uncertainty in the values of hadronic form factors at $q^2=0$.
 \begin{table}[htb]
\centering
\caption{Values of pole masses in GeV.} \label{Tab:polemass}
\begin{tabular}{|c|c|c|c|c|}
\hline
~current~ &~ $m(0^-)$~ &~ $m(0^+)$~ & ~$m(1^-)$~ &~ $m(1^+)$~\\ 
\hline
$\bar{u}b$ & 5.27 & 5.99 & 5.32 & 5.71\\
$\bar{c}b$ & 6.30 & 6.80 & 6.34 & 6.73\\
\hline
\end{tabular}
\end{table} 

\section{Constraints on new couplings}
In this analysis the new couplings are considered to be complex.  Considering the contribution of only  one  coefficient at a time  with  all others set to  zero, we  perform the chi-square fitting  for the individual complex  couplings. The $\chi^2$ is defined as 
\bea
\chi^2=\sum_i \frac{(\mathcal{O}_i^{\rm th}-\mathcal{O}_i^{\rm exp})^2}{(\Delta \mathcal{O}_i)^2},
\eea
where $\mathcal{O}_i^{\rm th}$ represent the theoretical predictions of the observables,  $\mathcal{O}_i^{\rm exp}$ symbolize the measured central values of the observables and $(\Delta \mathcal{O}_i)^2=(\Delta \mathcal{O}_i^{\rm th})^2+(\Delta \mathcal{O}_i^{\rm exp})^2$ contain  the  $1\sigma$ errors from theory and experiment. We constrain the real and imaginary parts of new coefficients related to $b \to c \tau \bar \nu_\tau$ quark level transitions from the $\chi^2$ fit of $R_{D^{(*)}}$, $R_{J/\psi}$ and Br($B_c^+ \to \tau^+  \nu_\tau$) observables and the couplings associated with $b \to u \tau \bar \nu_\tau$ processes are constrained from the fit of $R_\pi^l$, Br($B_u^+ \to \tau^+ \nu$) and Br($B^0 \to \pi^+ \tau^- \bar \nu$) data.  The updated values of all the observables  used for  fitting  are taken from \cite{Tanabashi:2018oca} and are listed in Table \ref{Tab:Fit}\,.   The  upper limit on the branching ratio of $B_c^+ \to \tau^+ \nu_\tau$ decay mode with the  present world average of the $B_c$ lifetime is \cite{Akeroyd:2017mhr}
\bea
{\rm Br}(B_c^+ \to \tau^+ \nu_\tau)\lesssim 30\%.
\eea 
We use the theoretical expressions of these observables and  their SM predictions  from \cite{Ray:2018hrx} and have listed them in Table \ref{Tab:Fit}\,.

\begin{table}[htb]
\centering
\caption{Values of the observables used in the fitting} \label{Tab:Fit}
\begin{tabular}{|c|c|c|}
\hline
Observables &~Experimental values~~&~SM Predictions~ \\
\hline
\hline
$R_D$ & $0.340 \pm 0.027 \pm 0.013$  &     $0.299 \pm 0.003$ \\
$R_{D^*}$ &  $ 0.295 \pm 0.011\pm0.008 $  &  $0.258 \pm 0.005 $\\
$R_{J/\psi}$  & $ 0.71 \pm 0.251 $   &$0.289 \pm 0.01$\\
${\rm Br}(B_c \to \tau \nu)$ & $< 30\%$     & $ (3.6 \pm 0.14) \times 10^{-2}$\\
\hline
$R_\pi^l$  & $0.699 \pm 0.156$  & $0.583\pm 0.055$\\
${\rm Br}(B_u \to \tau \nu)$ & $ (1.09\pm 0.24) \times 10^{-4}$ & $(8.48\pm0.5)  \times 10^{-5}$ \\
${\rm Br}(B^0 \to \pi^+ \tau \nu) $  & $< 2.5 \times 10^{-4}$  & $(9.40 \pm 0.75) \times 10^{-5}$\\
\hline
\end{tabular}
\end{table}

\begin{figure}
\includegraphics[scale=0.5]{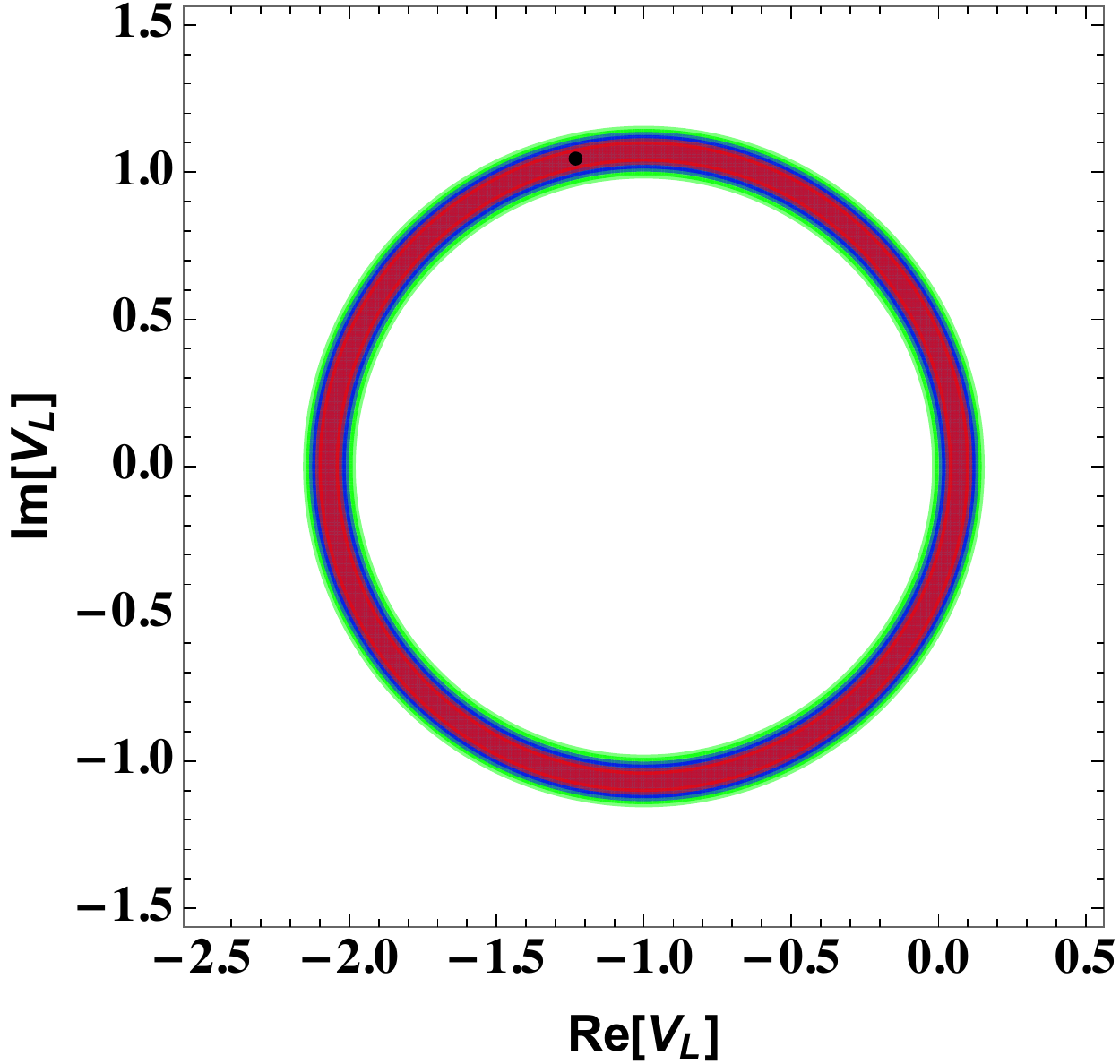}
\quad
\includegraphics[scale=0.5]{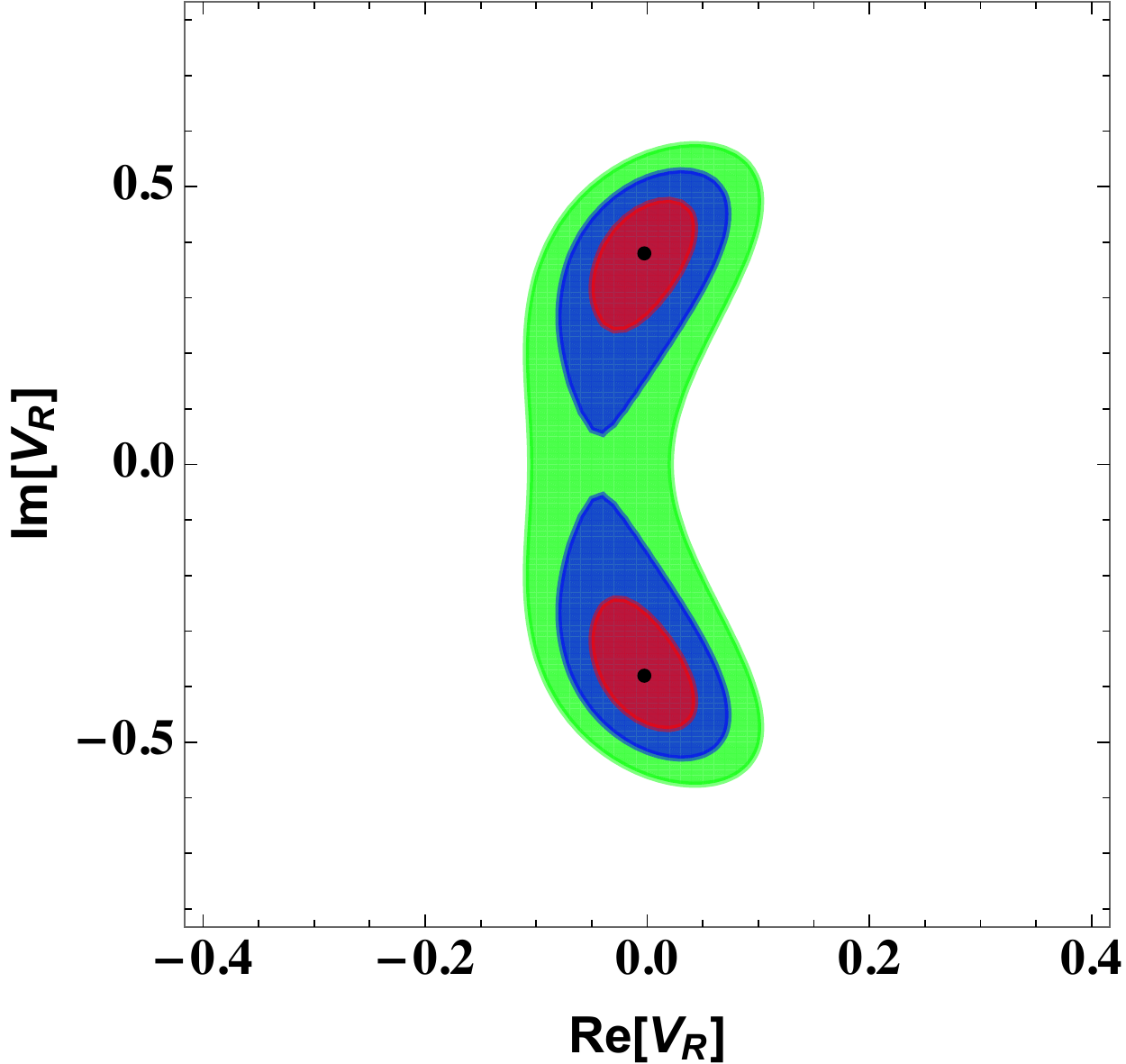}
\quad
\includegraphics[scale=0.5]{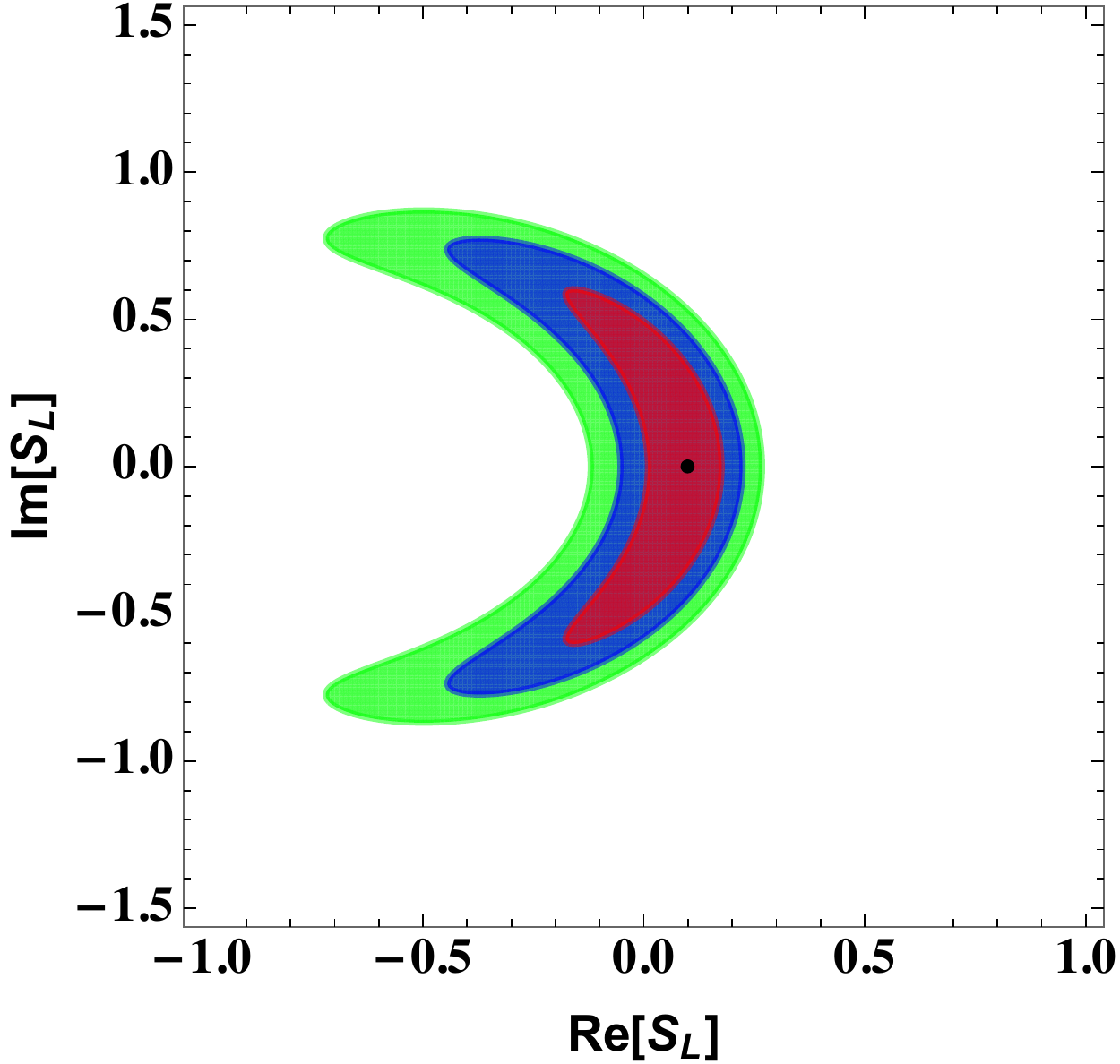}
\quad
\includegraphics[scale=0.5]{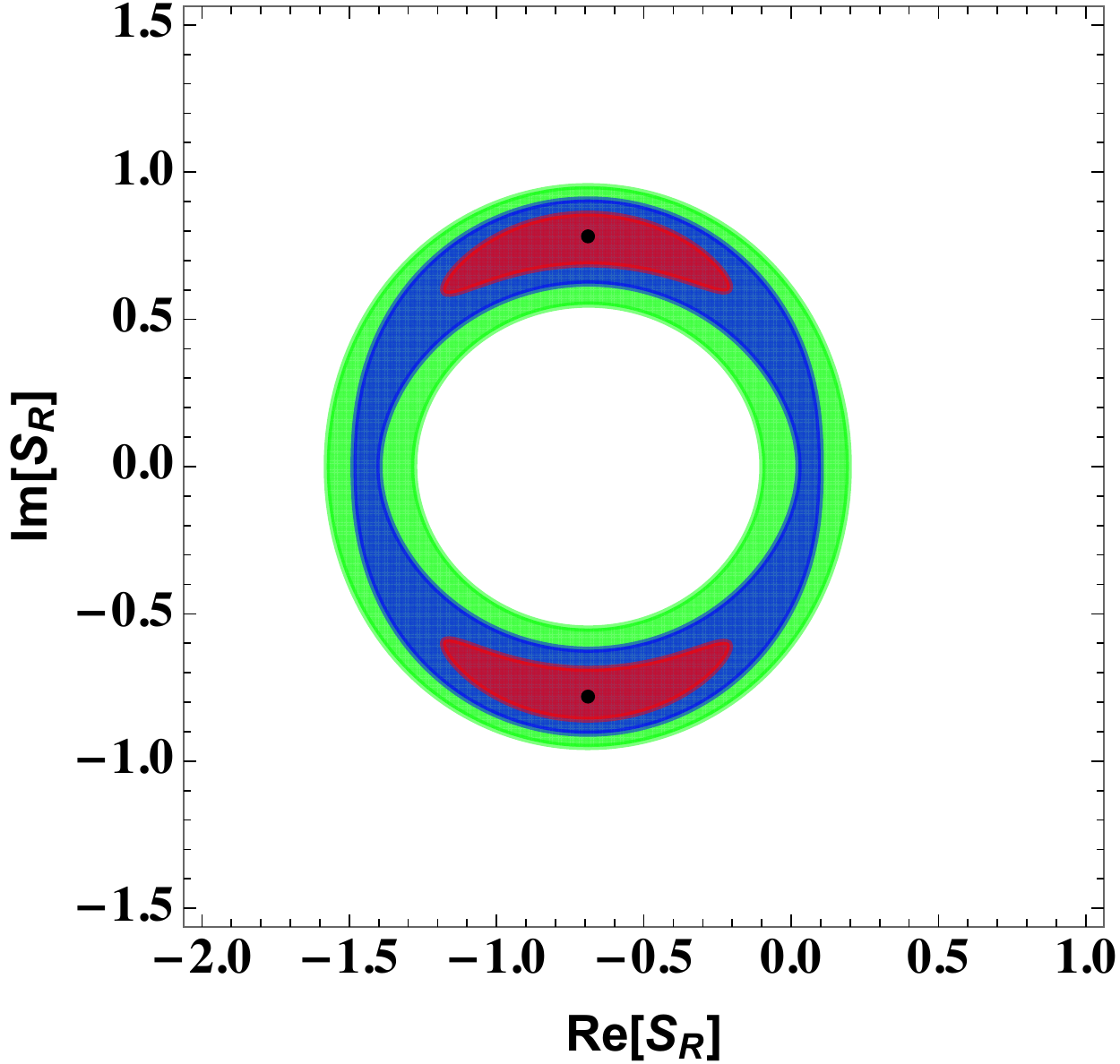}
\caption{Constraints on individual new complex  coefficients associated with $b \to c \tau \bar \nu_\tau$ processes from the $\chi^2$ fit of $R_{D^{(*)}}, R_{J/\psi}$ and upper limit on Br($B_c^+ \to \tau^+\nu_\tau$). Here the red, blue and green colors stand for   $1\sigma,~2\sigma$ and $3\sigma$ contours respectively. The black dots represent the best-fit values. }\label{Fig:bclnu-cntr}
\end{figure}

\begin{figure}
\includegraphics[scale=0.5]{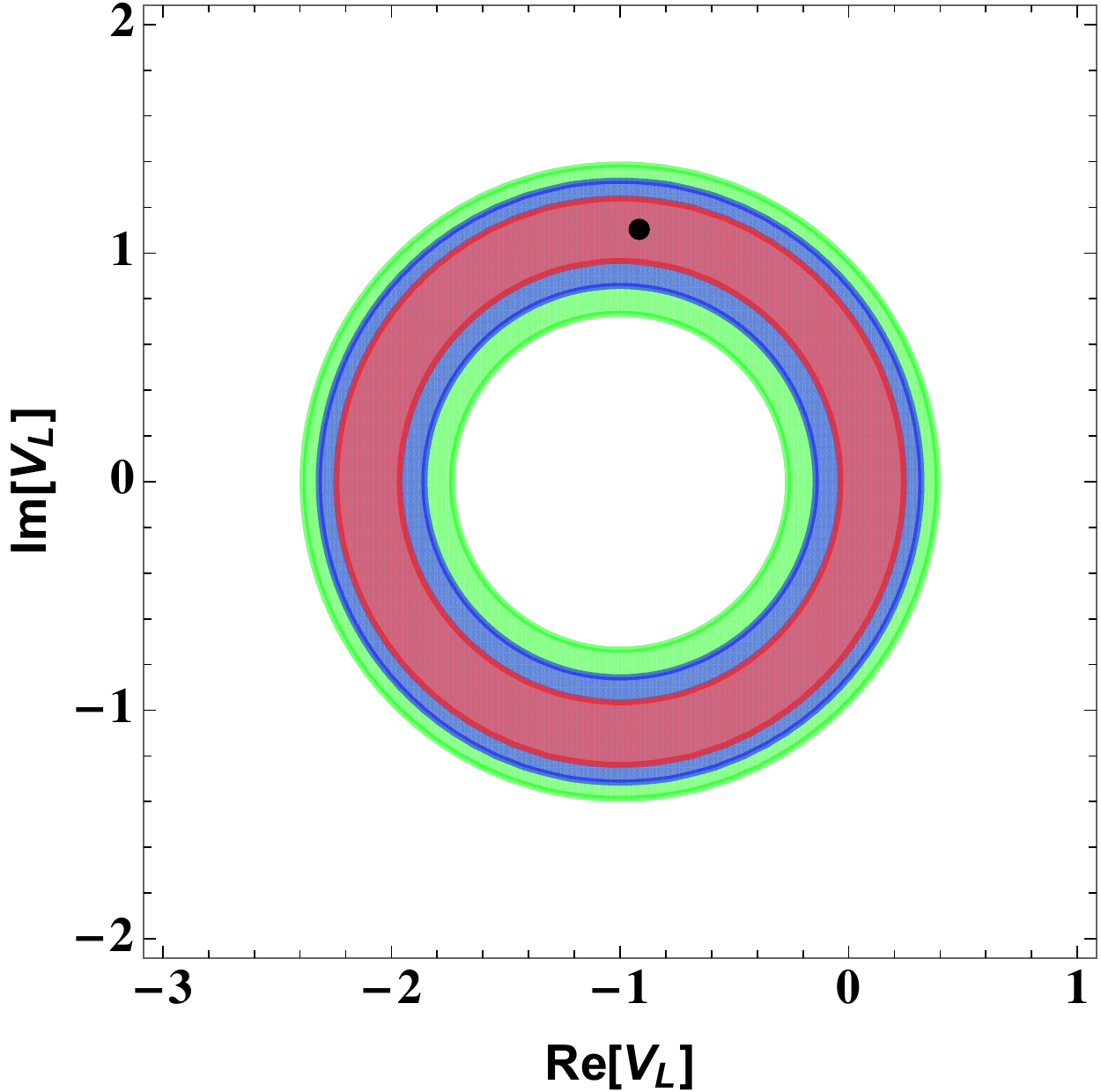}
\quad
\includegraphics[scale=0.5]{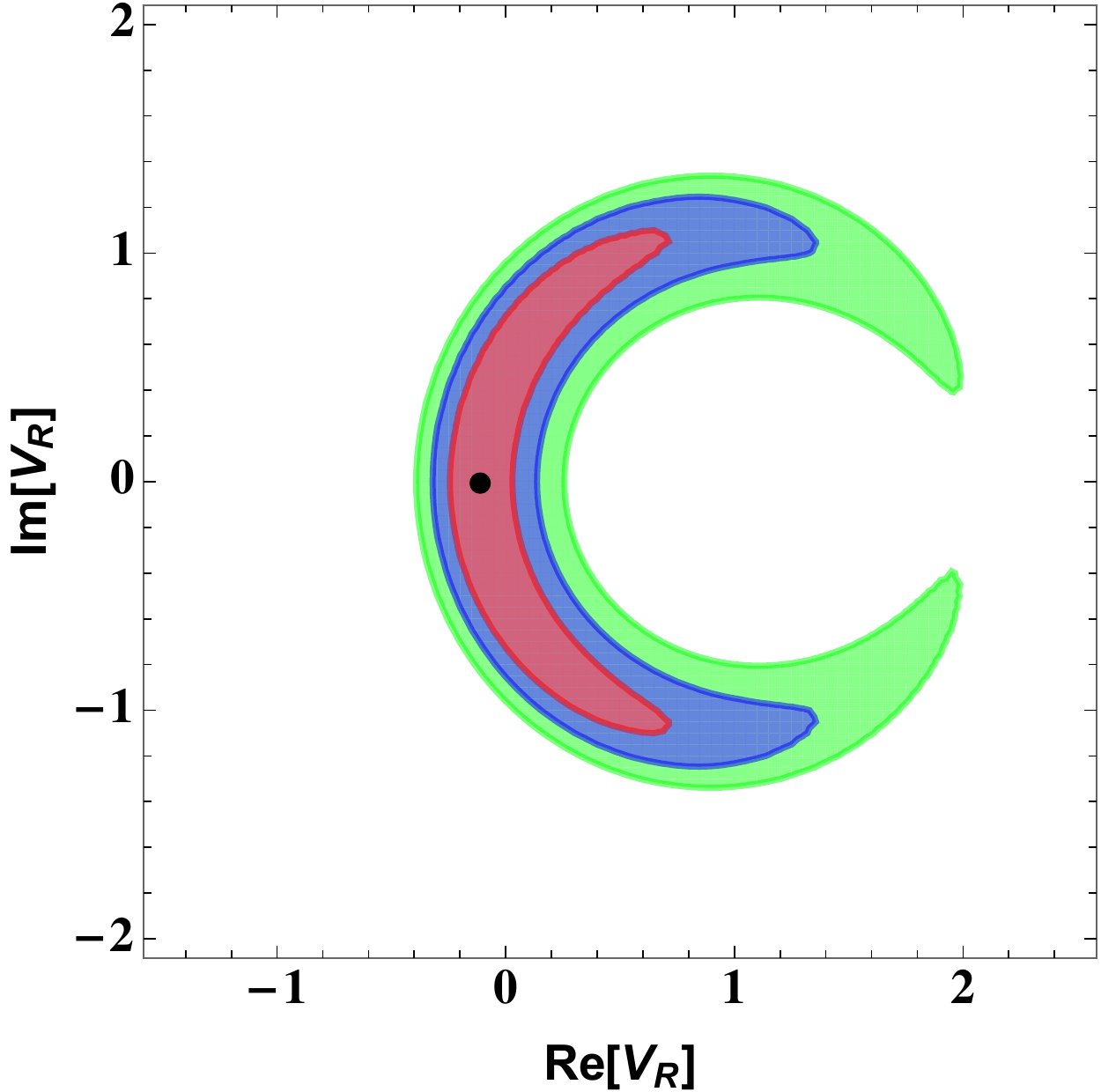}
\quad
\includegraphics[scale=0.5]{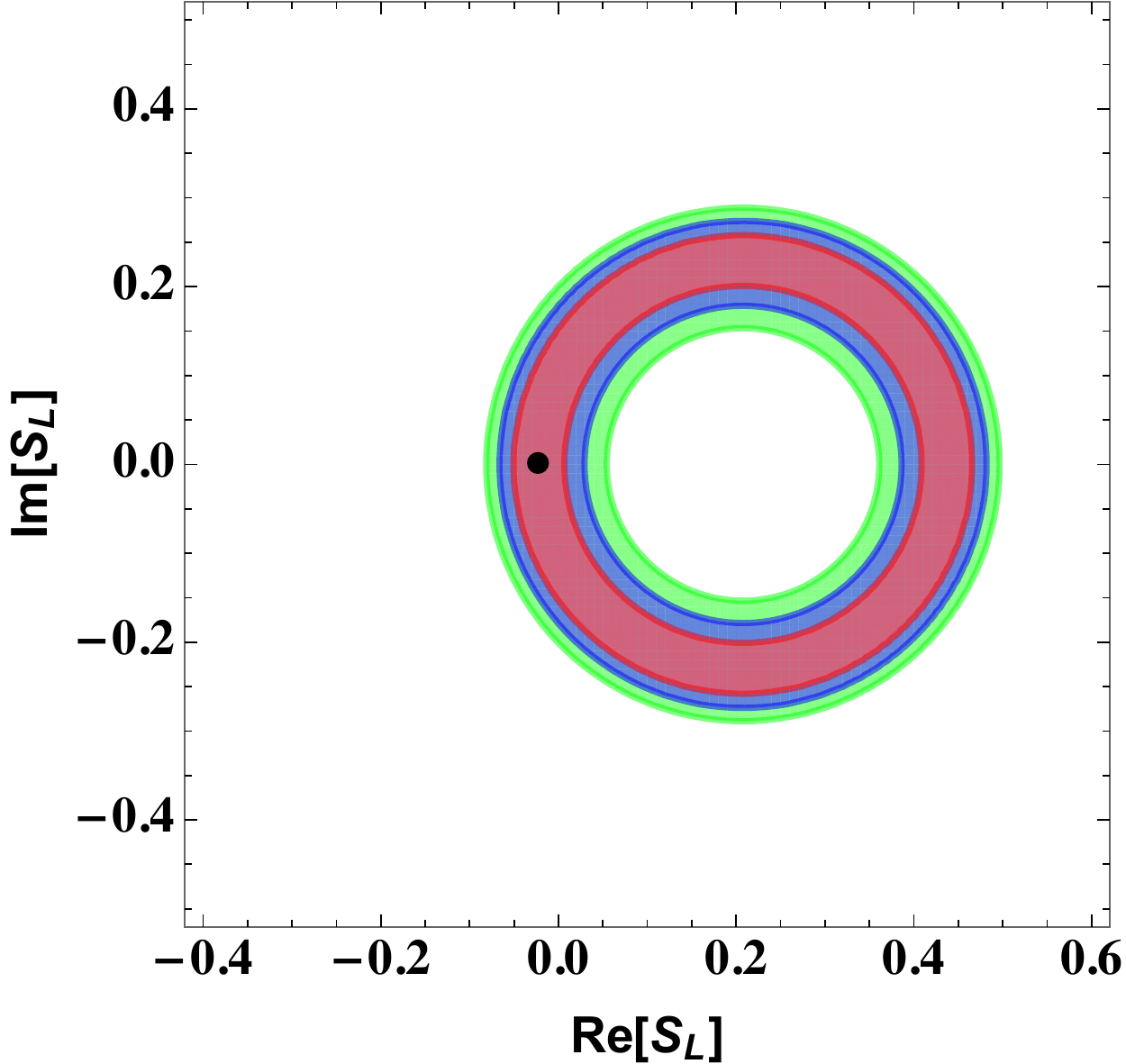}
\quad
\includegraphics[scale=0.5]{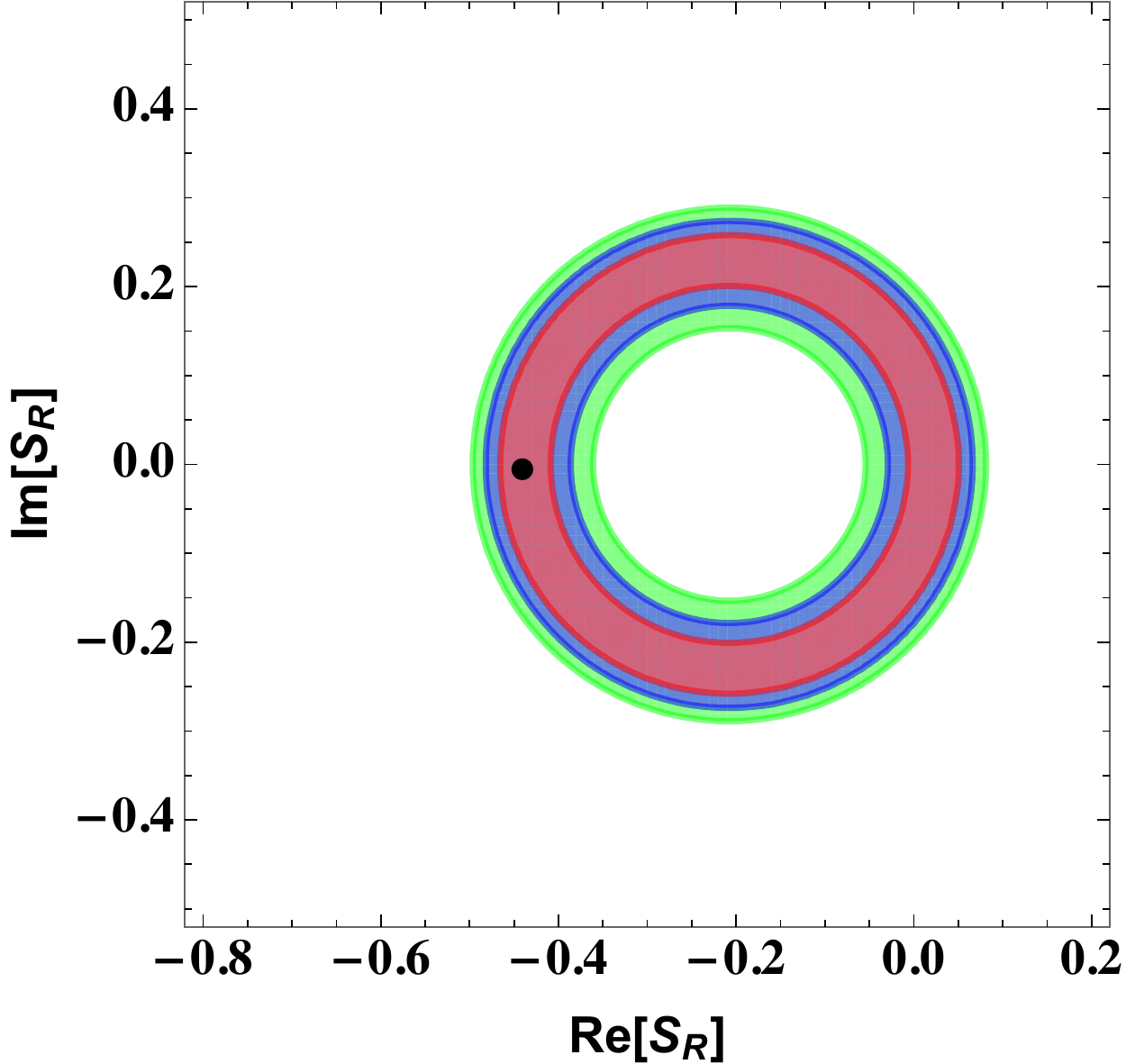}
\caption{Constraints on individual new complex  coefficients associated with $b \to u \tau \bar \nu_\tau$ processes from the $\chi^2$ fit of $R_{\pi}^l$,   Br($B_u^+ \to \tau^+\nu_\tau$) and upper limit on  Br($B^0 \to \pi^+ \tau^- \bar \nu_\tau$). }\label{Fig:bulnu-cntr}
\end{figure}

 In Fig. \ref{Fig:bclnu-cntr}\,, we present the constraints on $V_L$ (top-left panel), $V_R$ (top-right panel), $S_L$ (bottom-left panel) and $S_R$ (bottom-right panel) coefficients of $b \to c$ mediated decay modes and the corresponding plots for the coefficients of  $b \to u$ are shown in Fig. \ref{Fig:bulnu-cntr}\,. It should be noted that, the best-fit values are degenerate in the presence $V_L$ coupling ($V_L$, $S_L$ and $S_R$ couplings)  for $b \to c$  ($b \to u$) processes and for each of these couplings, we have considered only benchmark values.   The best-fit values and the corresponding $1\sigma$ ranges, which are obtained from the joint confidence regions of the real and imaginary planes of these  new couplings, are presented in Table \ref{Tab:con}\,. The $\chi^2/{\rm d.o.f}$, as well as the ${\rm pull} \simeq \sqrt{\chi^2_{\rm SM}-\chi^2_{\rm best-fit}}$, for all the coefficients are also listed in this Table. One can notice that, the Wilson coefficient corresponding to $b \to c $ scalar operators have $\chi^2/{\rm d.o.f}>1$, which implies that the fit is not robust. However, the pull values of $V_{L,R}$ coefficients of $b \to c$  implicit that the measured data are  consistent with our  model in the presence of either $V_L$ or $V_R$ and can be  a viable
candidate for explaining the $b \to c \tau \bar \nu_\tau$ anomalies.
 
\begin{table}[htb]
\centering
\caption{Best-fit values and corresponding  $1\sigma$  ranges (for one benchmark set only)   of the new complex coefficients.} \label{Tab:con}
\begin{tabular}{|c|c|c|c|c|c|}
\hline
Decay modes&New coefficients &Best-fit& $1\sigma$ range&~$\chi^2/{\rm d.o.f}$&Pull \\
\hline
\hline
$b \to c \tau \bar \nu_\tau $&$({\rm Re}[V_L], {\rm Im}[V_L]) $~&~
$(-1.233\;,1.045)$~&  $([-1.32,-1.075],[1.021,1.067])$ &~1.151~ &~$2.982$\\

&~$({\rm Re}[V_R], {\rm Im}[V_R]) $&$\left (-0.0034,~  -0.3783 \right )$~&~$\left ([-0.030,0.025],[-0.438,-0.31]\right )$&~$1.145$ &~$2.984$ \\

&~$({\rm Re}[S_L], {\rm Im}[S_L]) $&$\left (0.097\;,~ 0\right )$~&~$\left ([0.041,0.15],[-0.257,0.257] \right )$&~$4.213$ &~$1.663$ \\

&~$({\rm Re}[S_R], {\rm Im}[S_R]) $&$\left (-0.695, -0.777 \right )$~&~$\left ([-0.93,-0.55],[-0.835,-0.72]\right )$&~$2.175$ &~$2.616$ \\

\hline

$b \to u \tau \bar \nu_\tau$&$({\rm Re}[V_L], {\rm Im}[V_L]) $~&~$\left (-0.915,~ 1.108 \right )$~&~$([-1.45,-0.65],[1.02,1.19]) $~&~$0.131 $ &~$1.160$ \\

&~$({\rm Re}[V_R], {\rm Im}[V_R]) $&$(-0.116,  0)$~&~$([-0.205,-0.025], [-0.41,0.41])$&~$0.066$ &~$1.215$ \\

&~$({\rm Re}[S_L], {\rm Im}[S_L]) $&$(-0.024 ,0)$~&~$([-0.042,-0.004],[- 0.092,0.092])$&~$0.093$ &~$1.192$  \\

&~$({\rm Re}[S_R], {\rm Im}[S_R]) $&$\left (-0.439, 0.005   \right )$~&~$\left ([-0.457,-0.421] \;,[- 0.092,0.092]  \right )$&~$0.093$ &~$1.192$ \\
\hline
\end{tabular}
\end{table}
%
%
%
%
%
%
%
%

\section{Effect of new coefficients on $B_{d,s}^{*} \to (D,D_s,\pi,K) \tau \bar \nu_\tau$ decay modes}

After collecting all the theoretical expressions of required observables and getting knowledge on the allowed ranges of new parameters, we now proceed towards  numerical analysis. The particles masses and the values of the  CKM elements and the Fermi constant $G_F$ are  taken from PDG \cite{Tanabashi:2018oca}. The values of the current quark masses used in this analysis are as   $m_b=4.2 $ GeV,  $m_c=1.3 $ GeV,
and $m_u=2.2 $ MeV.  The $q^2$ dependence of the form factors,  required for numerical estimation are already discussed in section II. As the lifetimes of $B^*$ mesons are not yet measured, we impose the fact that for these mesons the electromagnetic transitions $B^* \to B\gamma $ are the dominant ones, and hence $\Gamma_{\rm tot}(B^*) \simeq \Gamma(B^* \to B \gamma)$ and use the following results \cite{Cheung:2014cka,Khodjamirian:2015dda}
\bea 
&&\Gamma(B_d^* \to B_d \gamma)=0.148 \pm 0.020 ~{\rm KeV}~~~~~~[34]\nn\\
&&\Gamma(B^{*+} \to B^{+} \gamma)=0.468_{-0.075}^{+0.073} ~{ \rm KeV}~~~~~~~[34]\nn\\
&&\Gamma(B_s^* \to B_s \gamma)\simeq 0.07 ~{\rm KeV}~~~~~~~~~~~~~~~~~~[35].\label{life-time}
\eea
From Eq. (\ref{life-time}), it should be noted that   $\Gamma_{\rm tot} (B^{*+} \simeq {1\over 3}\Gamma_{\rm tot} (B_d^{*}))$, so the branching fractions of $B^{*+} \to P \ell \nu_\ell$  processes are roughly one-third of $B_d^* \to P \ell \nu_\ell$. Hence, those results are not presented in this work.  Furthermore, we assume that the new physics will couple only to third generation leptons, so   the  $B_{d,s}^* \to P \mu \nu_\mu$ processes will not be affected by the presence of new physics operators, and their standard model branching fractions are listed in Table\;\ref{Tab:mu}\,, which are expected to be within the reach of LHC experiment.
 \begin{table}[htb]
\begin{center}
\caption{Branching fractions of  $B_{d,s}^* \to P \mu \bar{\nu}_\mu$ processes in the Standard Model. }\label{Tab:mu}
\begin{tabular}{| c |c |}
\hline
~Decay processes~ & ~SM Branching fraction~\\
 \hline
 \hline
~ $\rm{Br}(B^{*0} \to D^+ \mu^- \bar{\nu}_\mu)$ ~&$(9.318 \pm 1.901) \times 10^{-8}$  \\
  \hline
$\rm{Br}(B_s^{*} \to D_s^+ \mu^- \bar{\nu}_\mu)$ &$(1.709\pm0.349) \times 10^{-7}$ \\
 \hline 
  $\rm{Br}(B^{*0} \to \pi^+ \mu^- \bar{\nu}_\mu)$ &$(1.487 \pm 0.401)\times 10^{-9}$  \\
   \hline 
  $\rm{Br}(B_s^{*} \to K^+ \mu^- \bar{\nu}_\mu)$ &$(1.618\pm 0.437) \times 10^{-9}$ \\
 \hline
\end{tabular}
\end{center}
\end{table}

The $ \bar B_{d,s}^{*} \to (D,D_s) \tau^- \bar{\nu}_\tau$ processes  proceed through $b\to c$ quark level transitions, so we use the constrained values of the new couplings obtained for $b\to c \tau \bar{\nu}_\tau$ in order to calculate the associated observables of these processes. Similarly we use the allowed parameter space obtained for $b\to u \tau \bar{\nu}_\tau$ process to compute the observables associated with  $ B_{d,s}^{*} \to (\pi,K) \tau^- \bar{\nu}_\tau$ decay process as they are mediated by $b \to u$ quark level transitions. In the following subsections, we discuss the effect of the presence of one Wilson coefficient at a time on various observables of $B_{d,s}^{*} \to ( D,D_s,\pi,K) \tau \nu_\tau$ decay modes.

\subsection{Effect of $V_L$ only}
\begin{figure}
\includegraphics[scale=0.4]{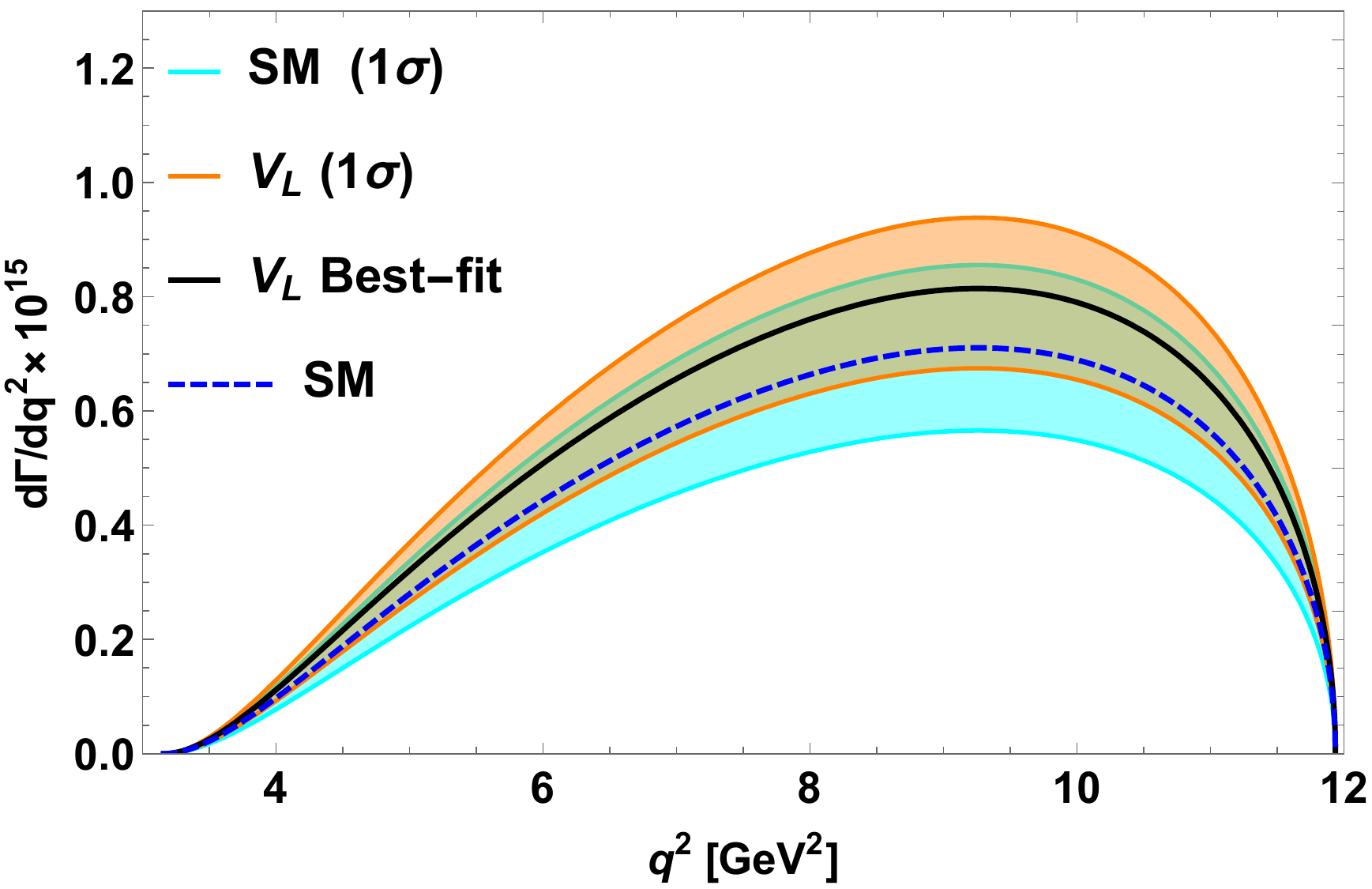}
\quad
\includegraphics[scale=0.4]{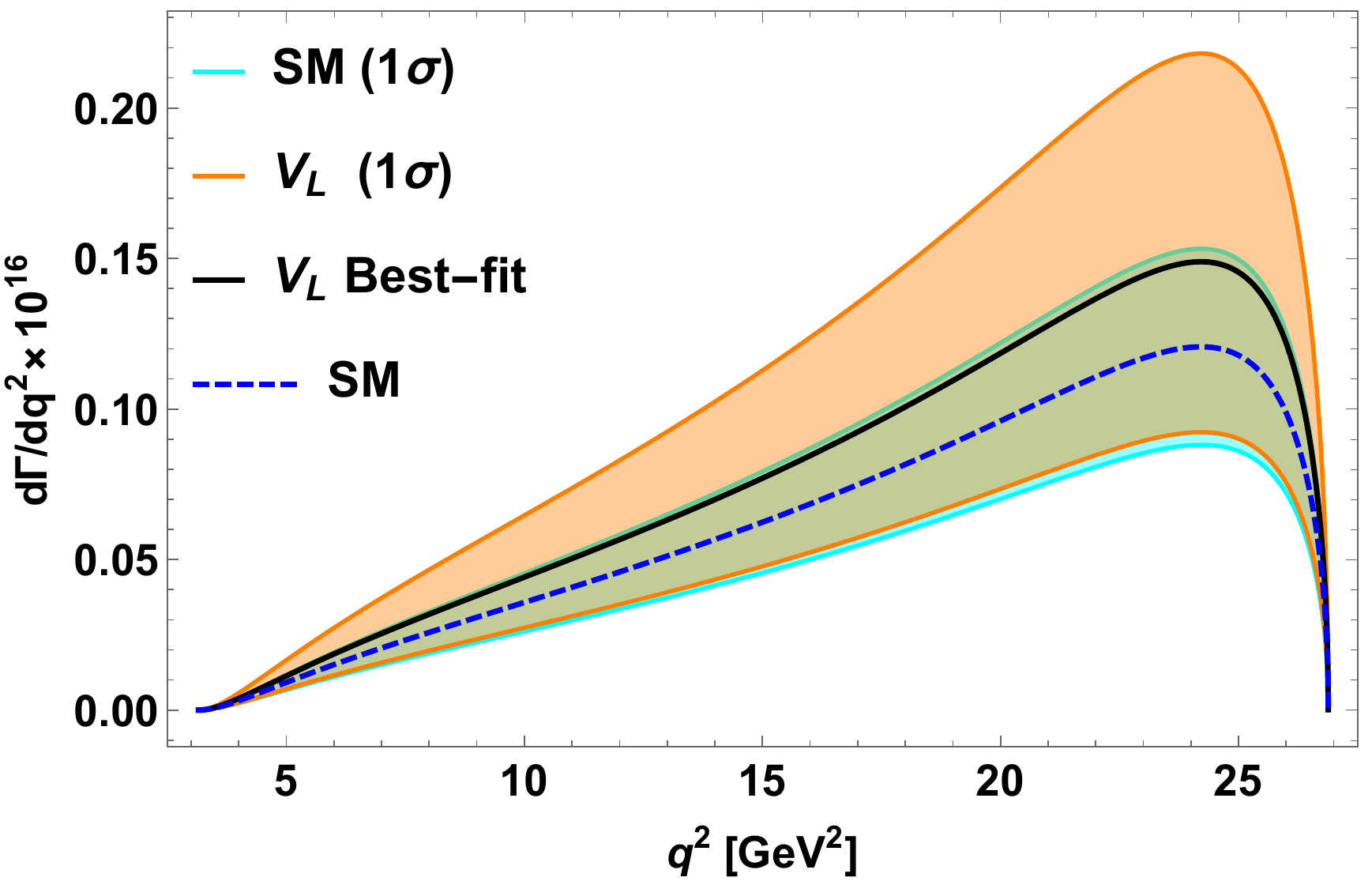}
\quad
\includegraphics[scale=0.4]{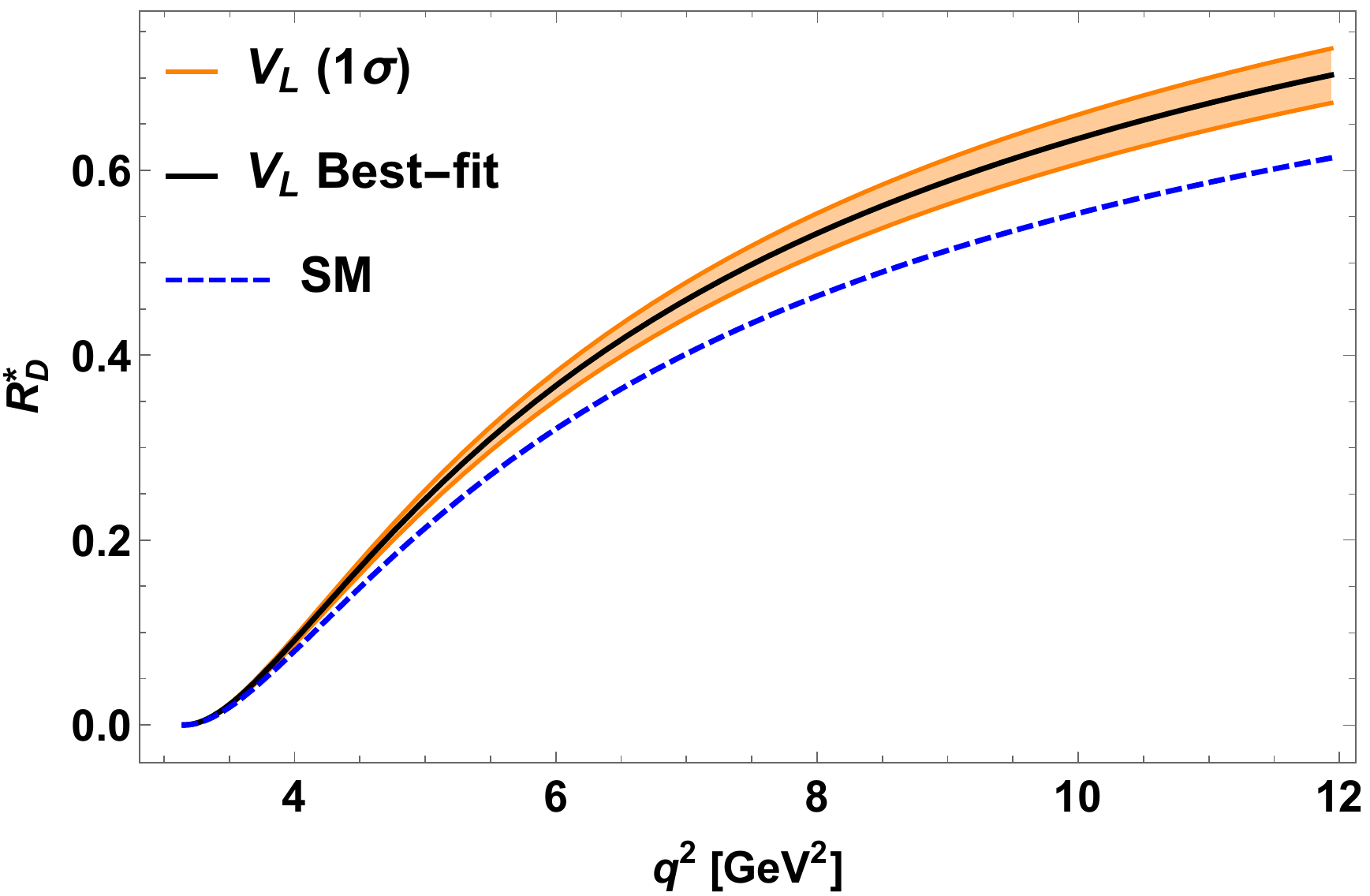}
\quad
\includegraphics[scale=0.4]{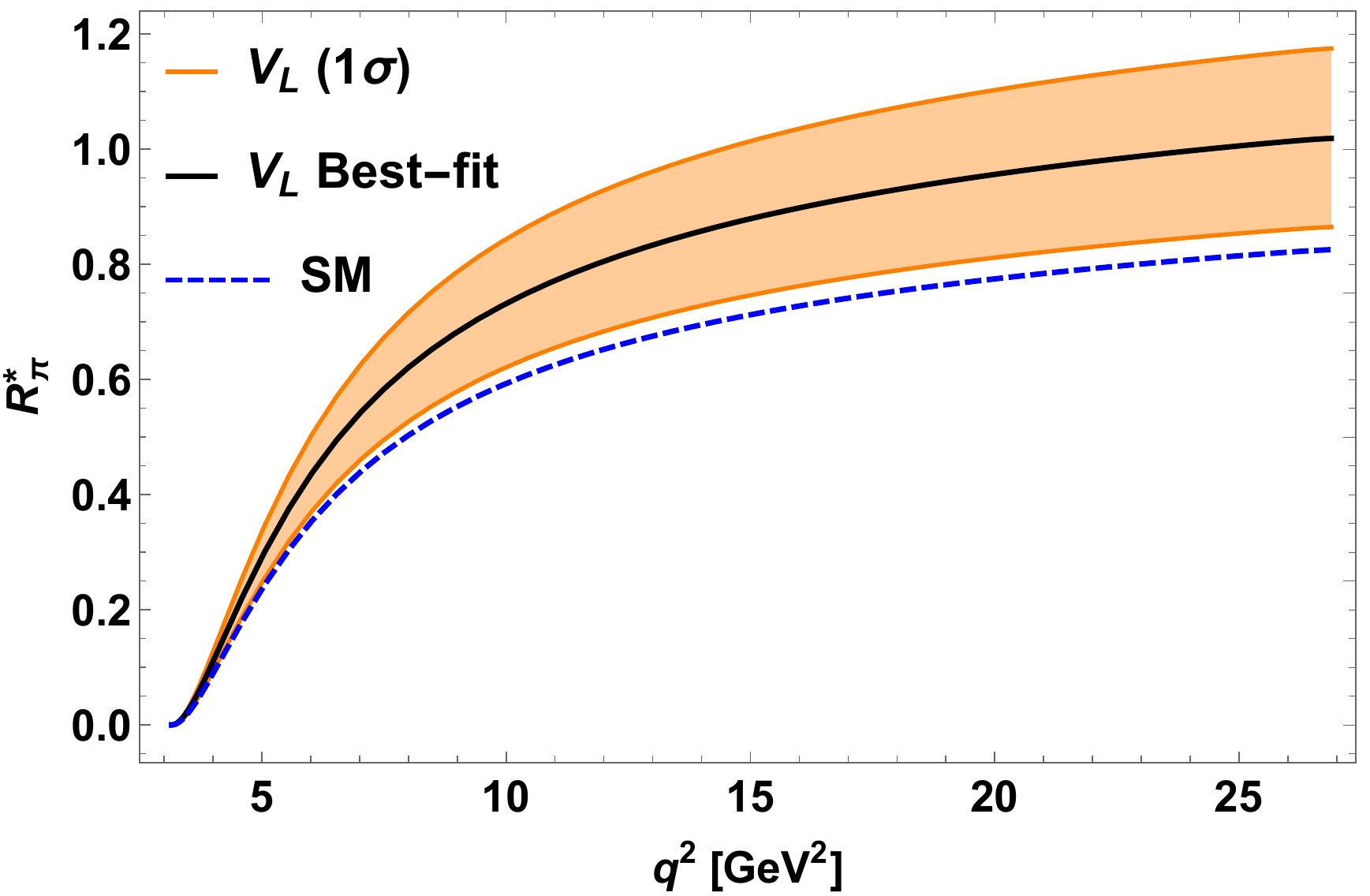}
\caption{The $q^2$ variation of differential decay rates and  LNU observables of $ \bar B_d^{*} \to D^+ \tau^- \bar{\nu_\tau}$ (left panel) and  $\bar{B}_d^* \to \pi^+ \tau \bar{\nu}_\tau$ (right panel) in presence of only $V_L$  new coefficient. Here the blue dashed lines represent the standard model predictions. The black solid lines and the orange bands are obtained by using the best-fit values and corresponding $1\sigma$ range of $V_L$ coefficient.  }
\label{variation-VL}
\end{figure}

Here we consider the case, where the  additional contribution to the SM Lagrangian  arising only  from $V_L$ coefficient  and  all other new coefficients are set to zero i.e., ($S_L=S_R=V_R=0$). Using the best-fit values and  $1\sigma$  allowed parameter space of $V_L$, obtained from the $\chi^2$ fit of $R_{D^{(*)}}, R_{J/\psi}$,  ${\rm Br}(B_c^+ \to \tau^+\nu)$  for $b \to c \tau \nu$ transitions  ($R_\pi^l,~ {\rm Br}(B^0 \to \pi^+ \tau^- \bar{\nu})$, ${\rm Br}(B_u^+ \to \tau^+\nu)$ for  $b \to u \tau \nu $ transitions), we then calculate the differential decay rate, LNU observable, lepton spin asymmetry and forward-backward asymmetry  of $B^{*0} \to D^+ \tau \nu$ and $B_s^* \to D_s^+ \tau \nu$ ($B^{*0} \to \pi^+ \tau \nu$ and  $B_s^* \to K^+ \tau \nu$) decay processes. In the left panel of Fig. \ref{variation-VL}\,, we show the $q^2$ variation of  decay rate (top) and $R_D^*$ observable (bottom) of $B^{* 0} \to D^+ \tau \nu$ process and the corresponding plots for $B^{*0} \to \pi^+ \tau \nu$ channel are presented in the right panel of this figure.  Here the blue dashed lines correspond to the SM prediction and the cyan bands represent the $1\sigma$ uncertainty, arising due to the errors in CKM matrix elements, hadronic form factors and the lifetime of $B^*$ meson.  The solid black lines are obtained by using the best-fit values of the left handed vectorial  new  $V_L$ coupling and the  orange bands represent  the  $1\sigma$ allowed ranges, which includes the SM uncertainties as well as the uncertainties due to the new couplings.  From the plots, one can notice  significant deviation in the branching ratios and LNU observables from their corresponding SM predictions due to presence of additional $V_L$ coefficient. 
To quantify these deviations, we define the pull metric at the  observable level as
\bea
{\rm Pull}_i=\frac{{\cal O}_i^{\rm NP} - {\cal O}_i^{\rm SM}}{\sqrt{{\Delta 
{\cal O}_i^{\rm NP}}^2 +\Delta {{\cal O}_i^{\rm SM}}^2}}\;,
\eea
where the index $i$ runs over all observables, ${\cal O}_i^{\rm SM}$ and  ${\cal O}_i^{\rm NP}$ denote the values of the observables  in SM and NP scenarios and   
${\Delta {\cal O}_i^{\rm SM}}$,  $\Delta {{\cal O}_i^{\rm NP}}$ are the corresponding $1\sigma$ uncertainties. We thus, obtain ${\rm Pull}_{\rm Br~(R_D^*)}=0.530 ~(4.0)$ for  $B^* \to D^+ \tau \nu$ process and  ${\rm Pull}_{\rm Br~(R_\pi^*)}=0.399 ~(1.239)$ for $B^* \to \pi \tau \nu$ process. The Pull value for $R_D^*$ and $R_\pi^*$ are found to be large as the SM uncertainties cancel out in these observables, thus providing significantly large pull value.
The plots for $B_s^* \to D_s^+ \tau \nu$ ($B_s^* \to K^+ \tau \nu$) process follow the same form as $B^{*0} \to D^+ \tau \nu$ ($B^{*0} \to \pi^+ \tau \nu$), and hence, are not included in this article.  The numerical values  of these observables are presented in Table \ref{Tab:VLVR}\,. Furthermore, no deviation has been observed in the forward-backward asymmetry and lepton-spin asymmetry observables from their  SM results,  so we don't provide the corresponding plots. The values of  $q^2$ at which the forward-backward asymmetry  vanishes are provided in Table \ref{Tab:zero-cross}\,.

\subsection{Effect of $V_R$ only}

\begin{figure}
\includegraphics[scale=0.4]{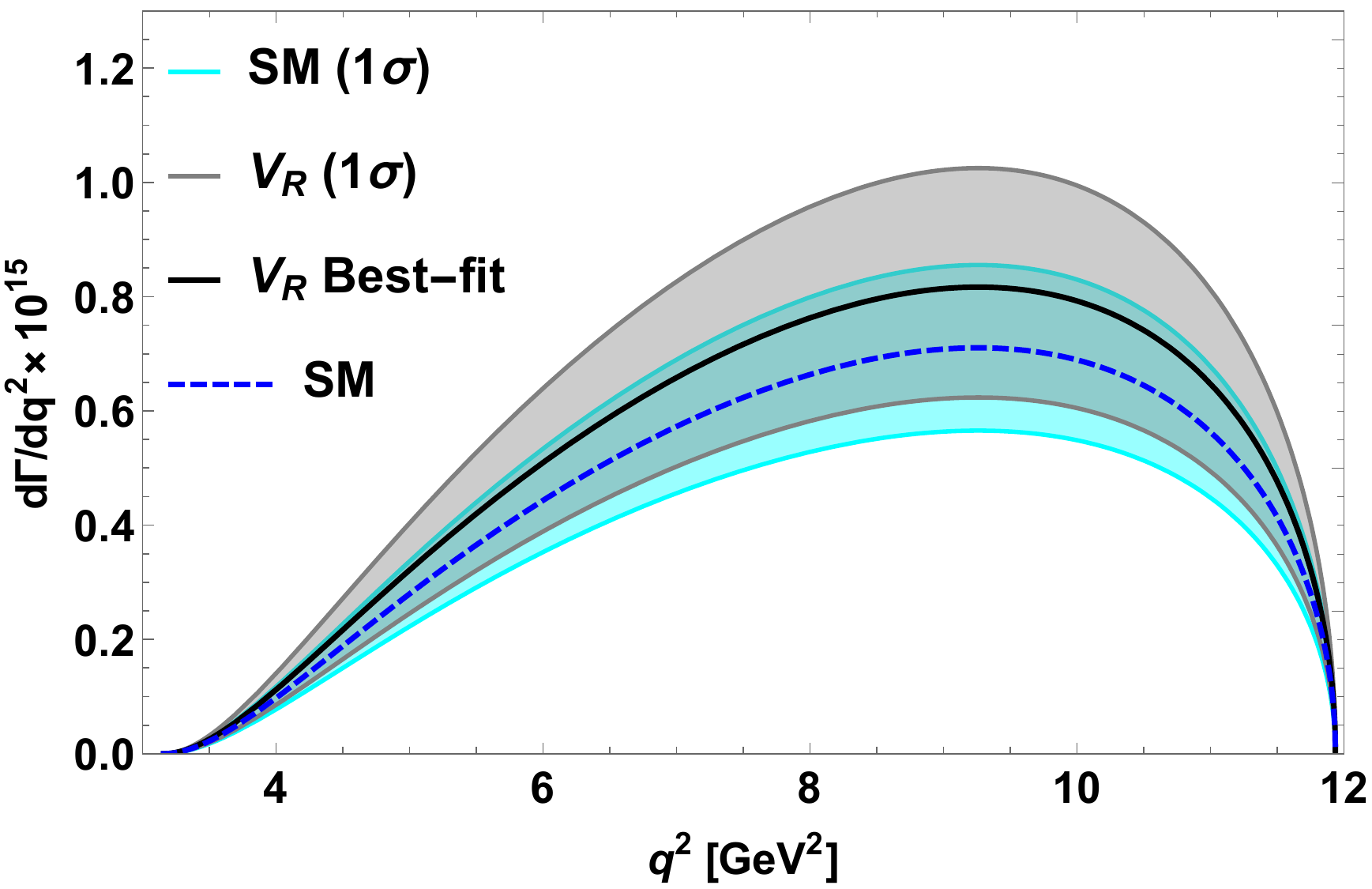}
\quad
\includegraphics[scale=0.4]{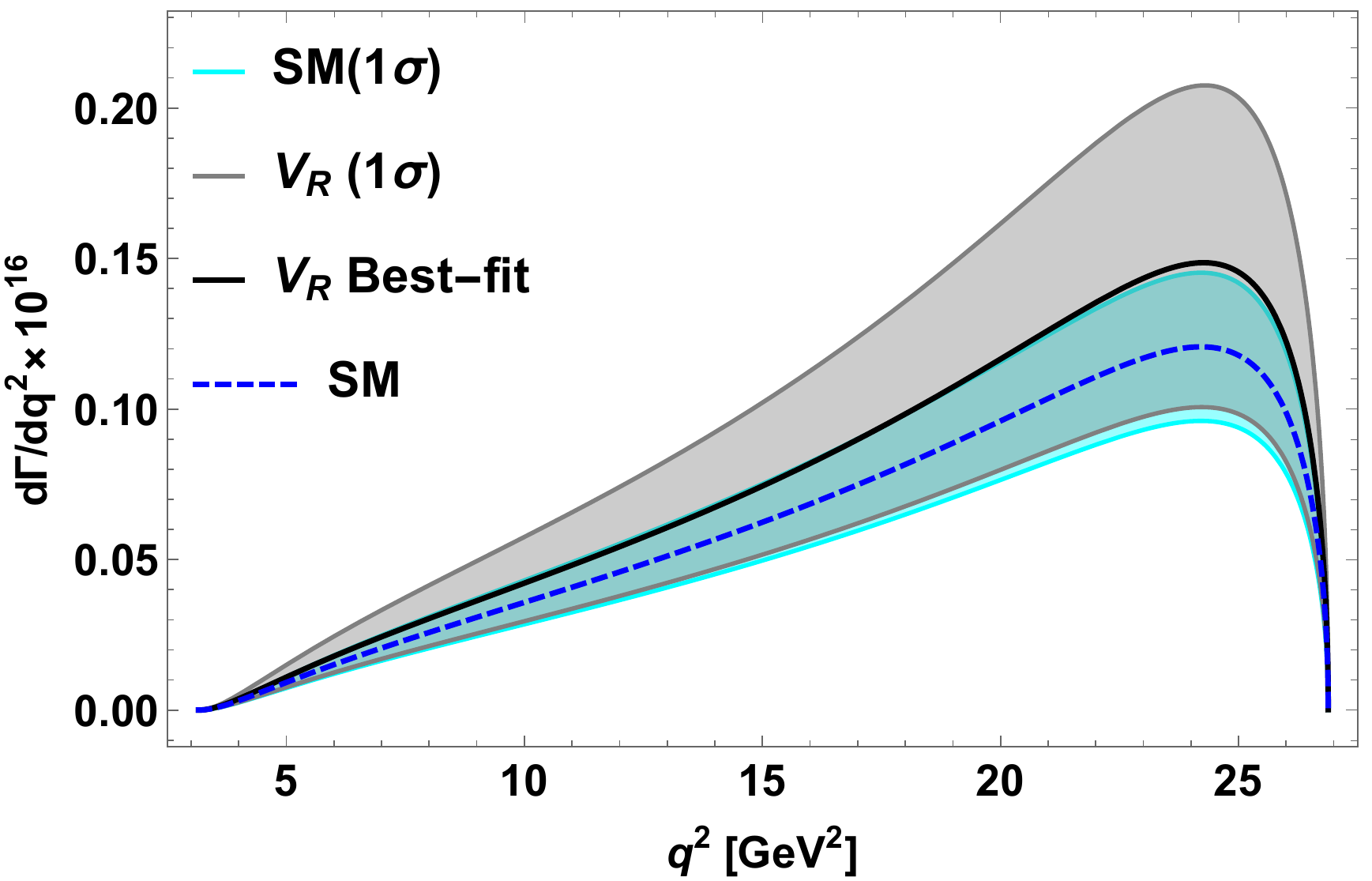}
\quad
\includegraphics[scale=0.4]{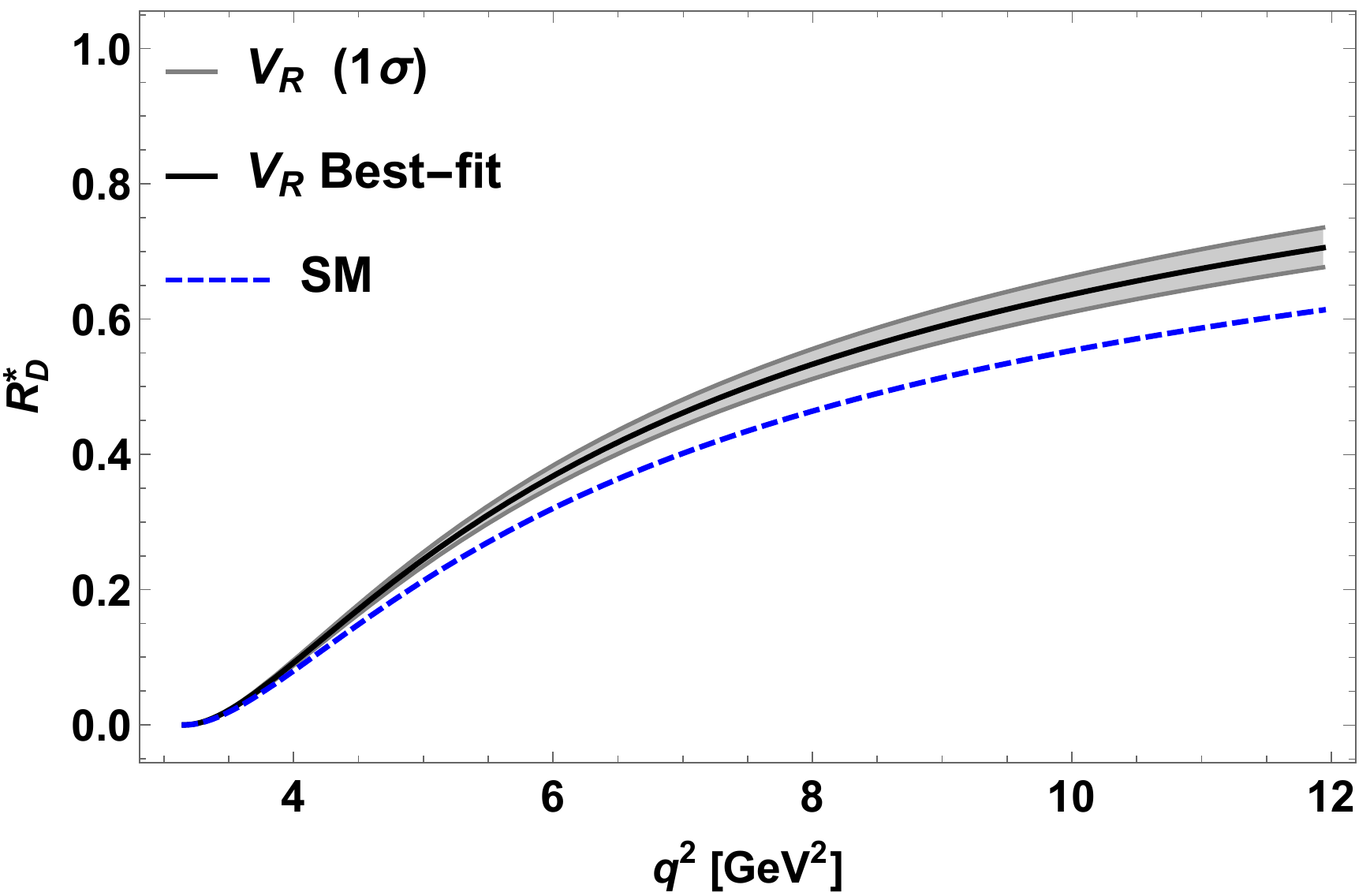}
\quad
\includegraphics[scale=0.4]{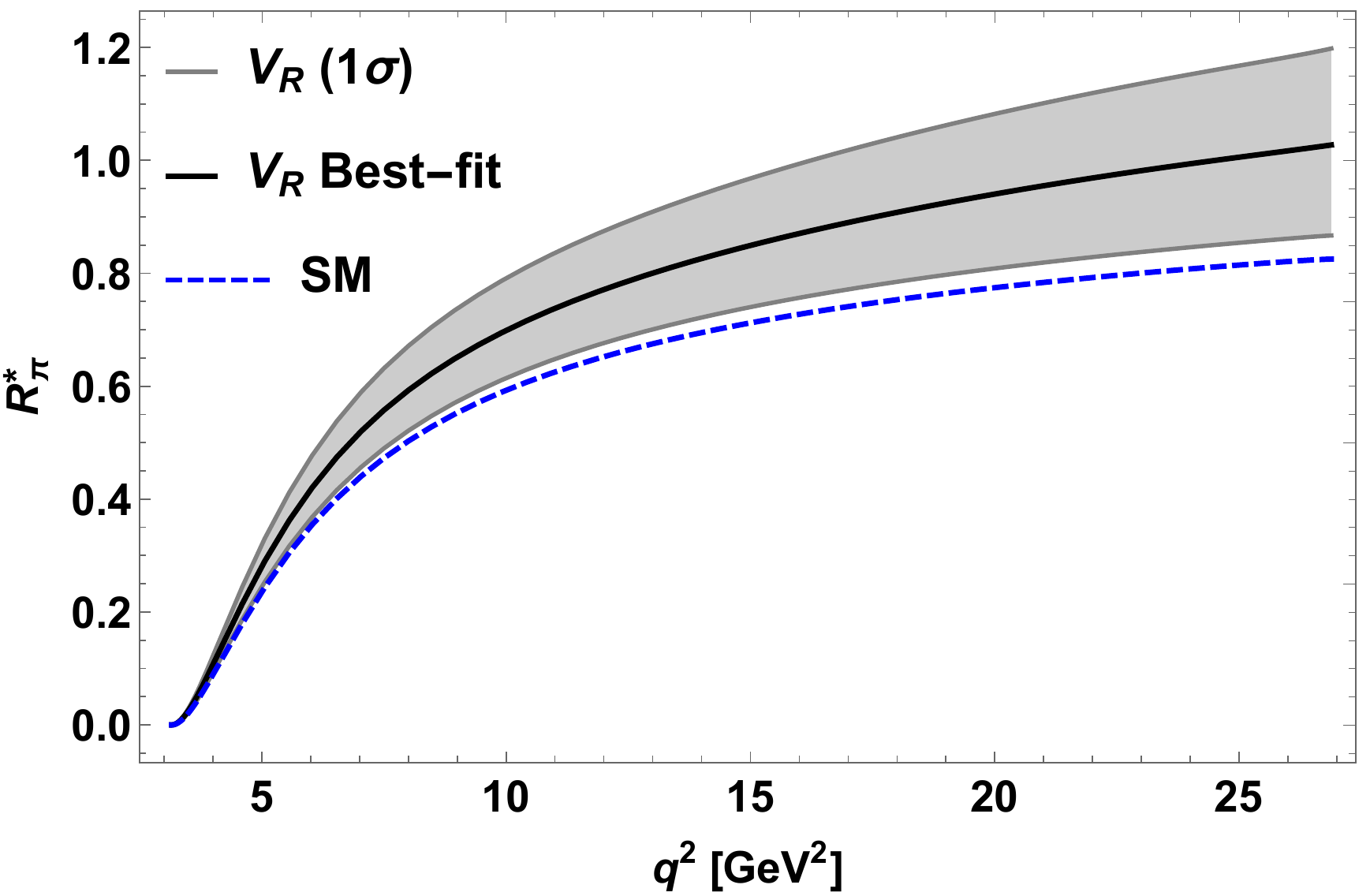}
\quad
\includegraphics[scale=0.4]{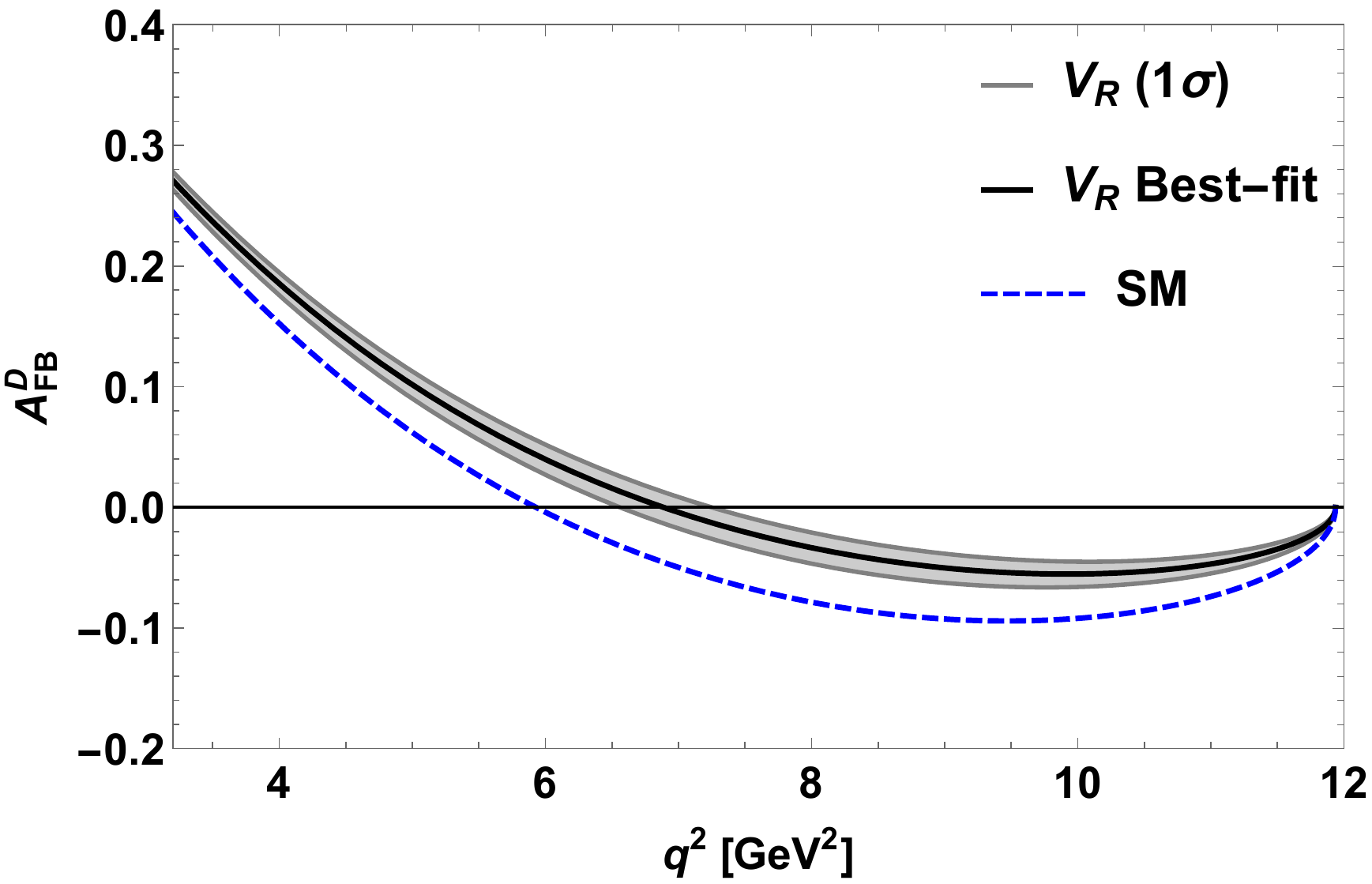}
\quad
\includegraphics[scale=0.4]{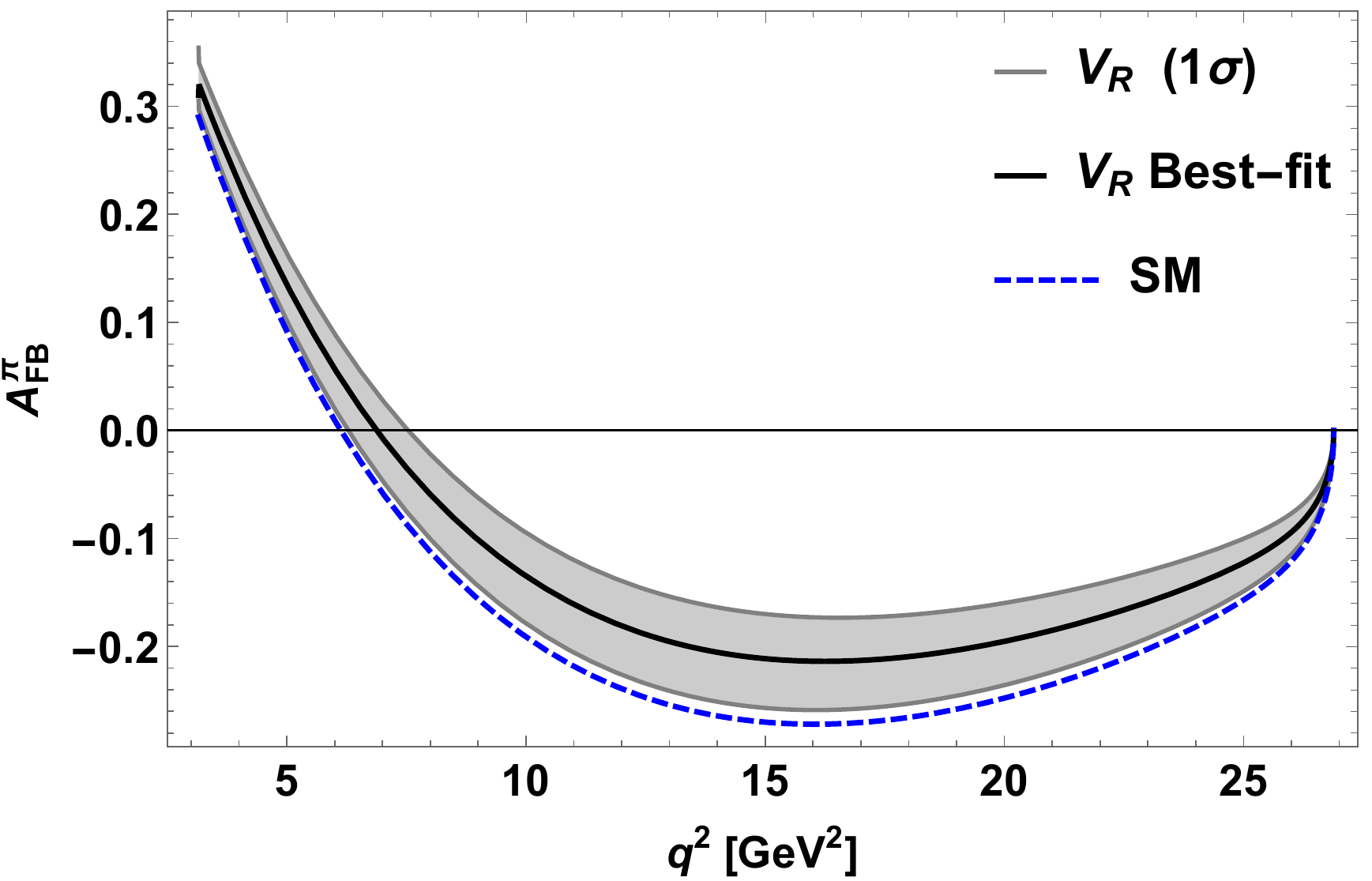}

\caption{The $q^2$ variation of differential decay rate, lepton nonuniversaity parameter and forward-backward asymmetry of $ \bar B_d^{*} \to D^+ \tau^- \bar{\nu}$ (left panel) and  $\bar{B}_d^* \to \pi^+  \tau \bar{\nu}$ (right panel) in presence of new $V_R$ coefficient. The black solid lines and the gray bands are obtained by using the best-fit values and corresponding $1\sigma$ range of $V_R$ coefficient. }.
\label{variation-VR}
\end{figure}

In this scenario, we explore the effect of only $V_R$ coefficient on the decay rate and angular observables of $B^* \to (D^+,\pi^+) \tau \nu_\tau$ processes. Using the best-fit values and corresponding $1\sigma$ allowed ranges of $V_R$ coefficients associated with $b \to (c,u)\tau \bar \nu_\tau$ transitions, we present the plots for the decay rate (left-top panel), $R_{D}^*$  (left-middle) and forward-backward asymmetry (left-bottom panel) of $B^* \to D^+ \tau \nu$ decay modes in Fig. \ref{variation-VR}\,. The corresponding plots for $B^* \to \pi^+ \tau \nu$ process are depicted in the right panel of Fig. \ref{variation-VR}\,. Here the solid black lines  are obtained by using the best-fit values of new $V_R$ couplings and the gray bands   by including  $1\sigma$ uncertainties of all input values. Reasonable deviation in all the observables (except the lepton-spin asymmetry)  from their SM results are  found due to the presence of additional $V_R$ coefficient, with Pull values ${\rm Pull}_{{\rm Br}/{R_D^*}/{\rm A_{FB}}}=0.429/3.21/3.391$ for $B^* \to D^+ \tau \nu$ process and ${\rm Pull}_{{\rm Br}/{R_\pi^*}/{\rm A_{FB}}}=0.368/1.203/1.323$ for $B^* \to \pi \tau \nu$. In Table \ref{Tab:VLVR}\,, we present the numerical values of decay rates and  all these parameters.  Due to the additional contribution from $V_R$ coefficient, we notice deviation in the zero crossing of the forward-backward asymmetry towards high $q^2$ and the $q^2$ values of the zero crossing point are given in Table \ref{Tab:zero-cross}\,.

\begin{table}[htb]
\begin{center}
\caption{Predicted numerical values of differential decay rate,  LNU observables, lepton spin asymmetry and forward-backward asymmetry of $ \bar B_{d,(s)}^{*} \to D^+(D_s^+) \tau^- \bar{\nu}_\tau$ and  $\bar{B}_{d(s)}^* \to \pi^+(K^+) \tau \bar{\nu}_\tau$ decay processes in the SM and in the presence of $V_{L,R}$ coefficients. }\label{Tab:VLVR}
\begin{tabular}{| c |c | c |  c|}
\hline
Observables~ & ~SM Predictions~ & Values with  $V_L$ & Values with $V_R$   \\
 \hline
 \hline
 $\rm{Br}(B^{*0} \to D^+ \tau^- \bar{\nu}_\tau)$ &$(2.786 \pm 0.568)\times 10^{-8}$  &
 $[2.646, 3.679] \times 10^{-8}$&   $[2.444,4.019] \times 10^{-8}$\\
  $R^*_D$ & $0.299$ & $[0.328, 0.357] $ & $[0.330, 0.358]$ \\
 $A_\lambda^D$ & $0.576$ & $0.576$ & $ 0.576$\\
$A_{\rm FB}^D$ & $-0.054$ & $-0.054$ & $[-0.027, -0.004]$\\  
 \hline
$\rm{Br}(B_s^{*0} \to D_s^+ \tau^- \bar{\nu}_\tau)$ &$(5.074\pm1.035) \times 10^{-8}$  &
 $[4.818,6.701] \times 10^{-8}$&   $[4.453, 7.320] \times 10^{-8}$\\
  $R^*_{D_s}$ & $0.297$ & $[0.326,0.354] $ & $[0.327, 0.356]$ \\
  $A_\lambda^{D_s}$ & $0.573$ & $0.573$ & $0.573$\\
$A_{\rm FB}^{D_s}$ & $-0.053$ & $-0.053$ & $[-0.025,-0.003]$\\  
 \hline 
 
  $\rm{Br}(B^{*0} \to \pi^+ \tau^- \bar{\nu}_\tau)$ &$(1.008\pm 0.272) \times 10^{-9}$  &
 $[0.771,1.821]  \times 10^{-9}$&   $(0.767, 1.781) \times 10^{-9}$\\
 $R^*_\pi$ &0.678 & $[0.710, 0.965]$ & $[0.707, 0.943]$ \\
 $A_\lambda^{\pi}$ & 0.781 & 0.781 & $[0.780,  0.781]$\\
$A_{\rm FB}^\pi$ & $-0.209$ & $-0.209$ & $[-0.198, -0.127]$\\  
 \hline 
 
  $\rm{Br}(B_s^{*0} \to K^+ \tau^- \bar{\nu}_\tau)$ &$(1.034 \pm 0.279) \times 10^{-9}$  &
 $[0.791, 1.869]  \times 10^{-9}$&   $[0.787,1.818]  \times 10^{-9}$\\
 
 $R^*_K$ &0.639 & $[0.670, 0.910]$ & $[0.666, 0.885]$ \\
 $A_\lambda^{K}$ & 0.747 & 0.747 & $[0.745, 0.746]$\\
$A_{\rm FB}^K$ & $-0.207$ & $-0.207$ & $[-0.196 , -0.123]$\\ 
 \hline
\end{tabular}
\end{center}
\end{table}

\subsection{Effect of $S_L$ only}

\begin{figure}
\includegraphics[scale=0.4]{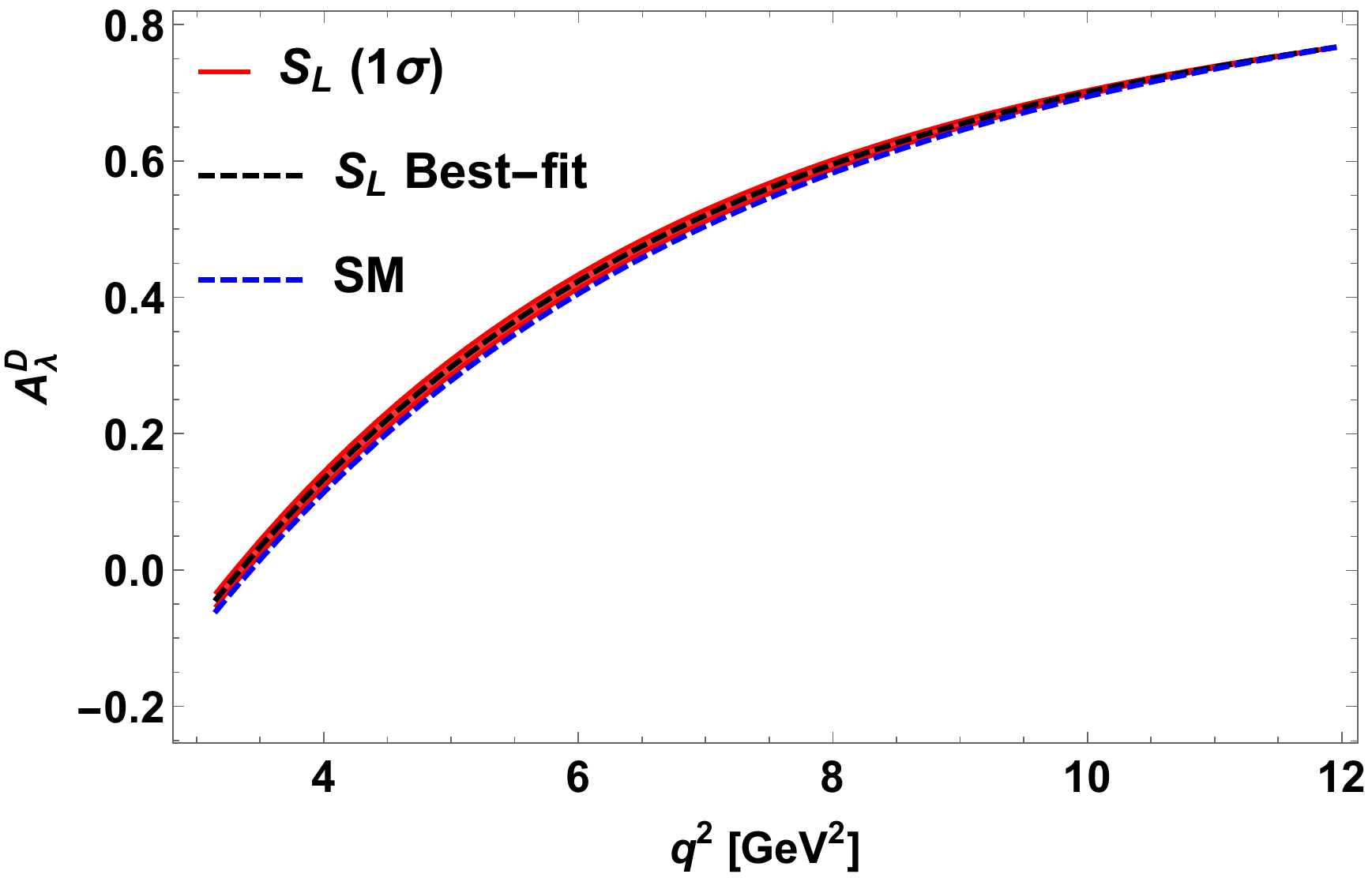}
\quad
\includegraphics[scale=0.4]{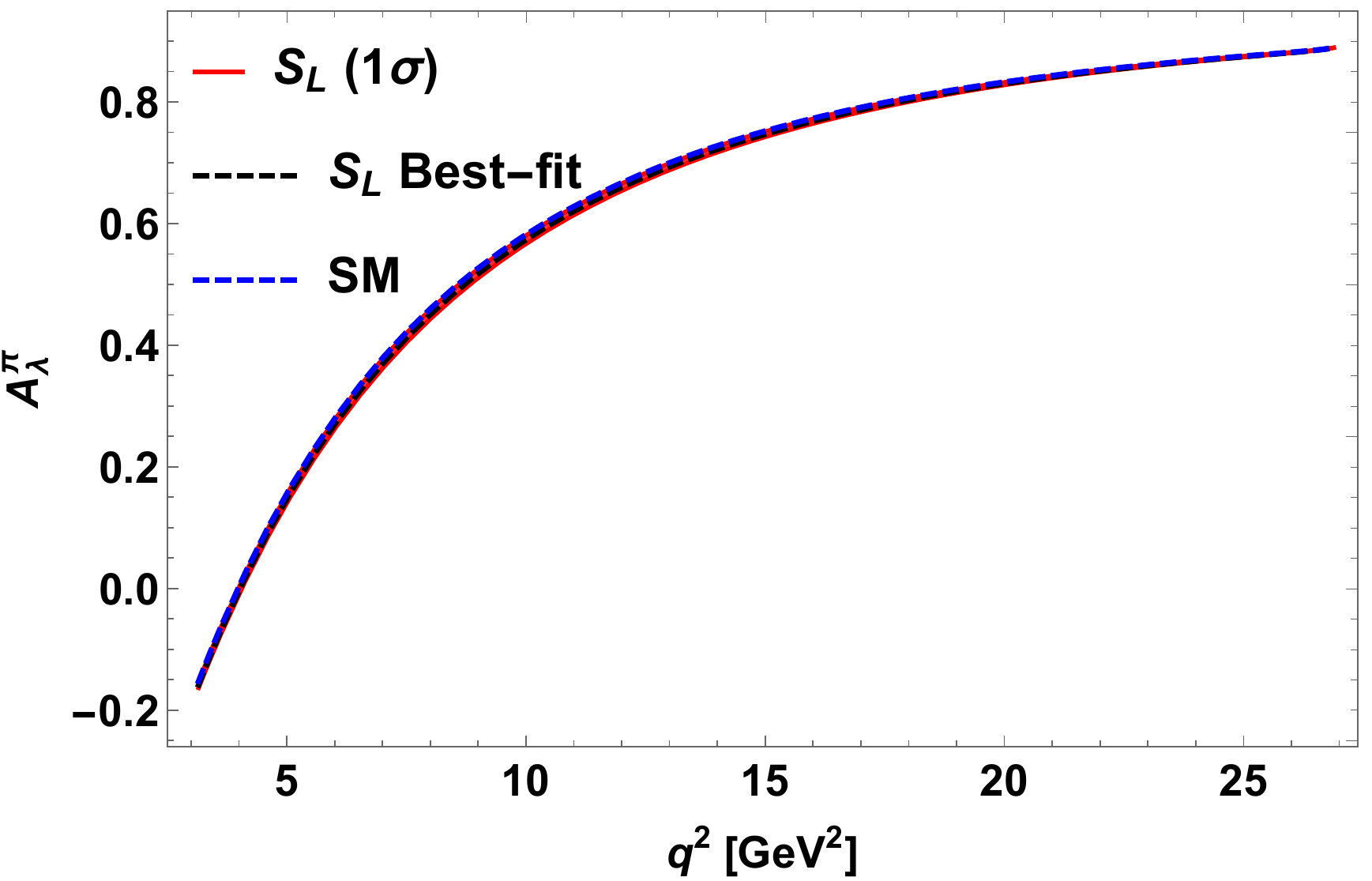}
\quad
\includegraphics[scale=0.4]{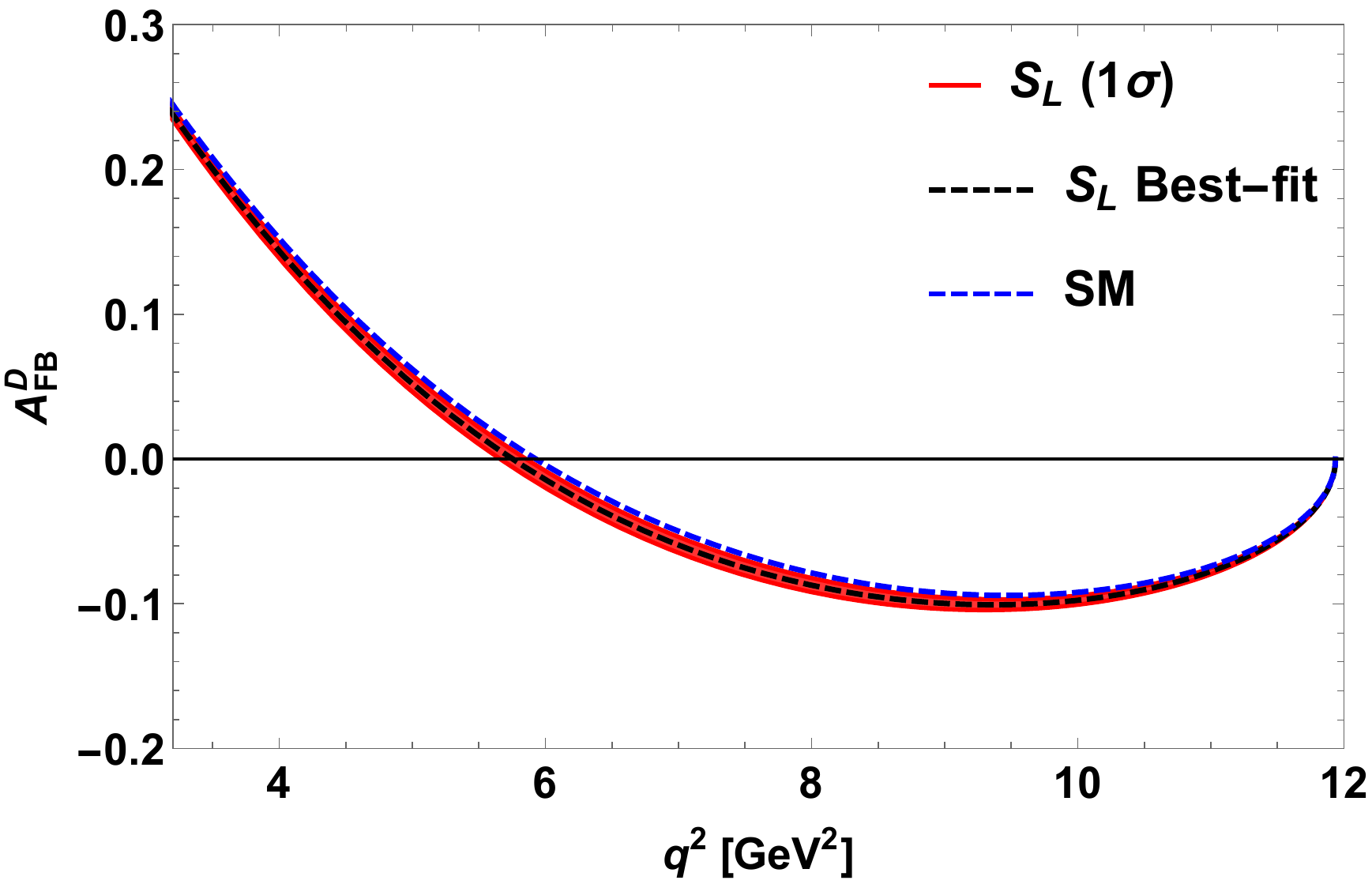}
\quad
\includegraphics[scale=0.4]{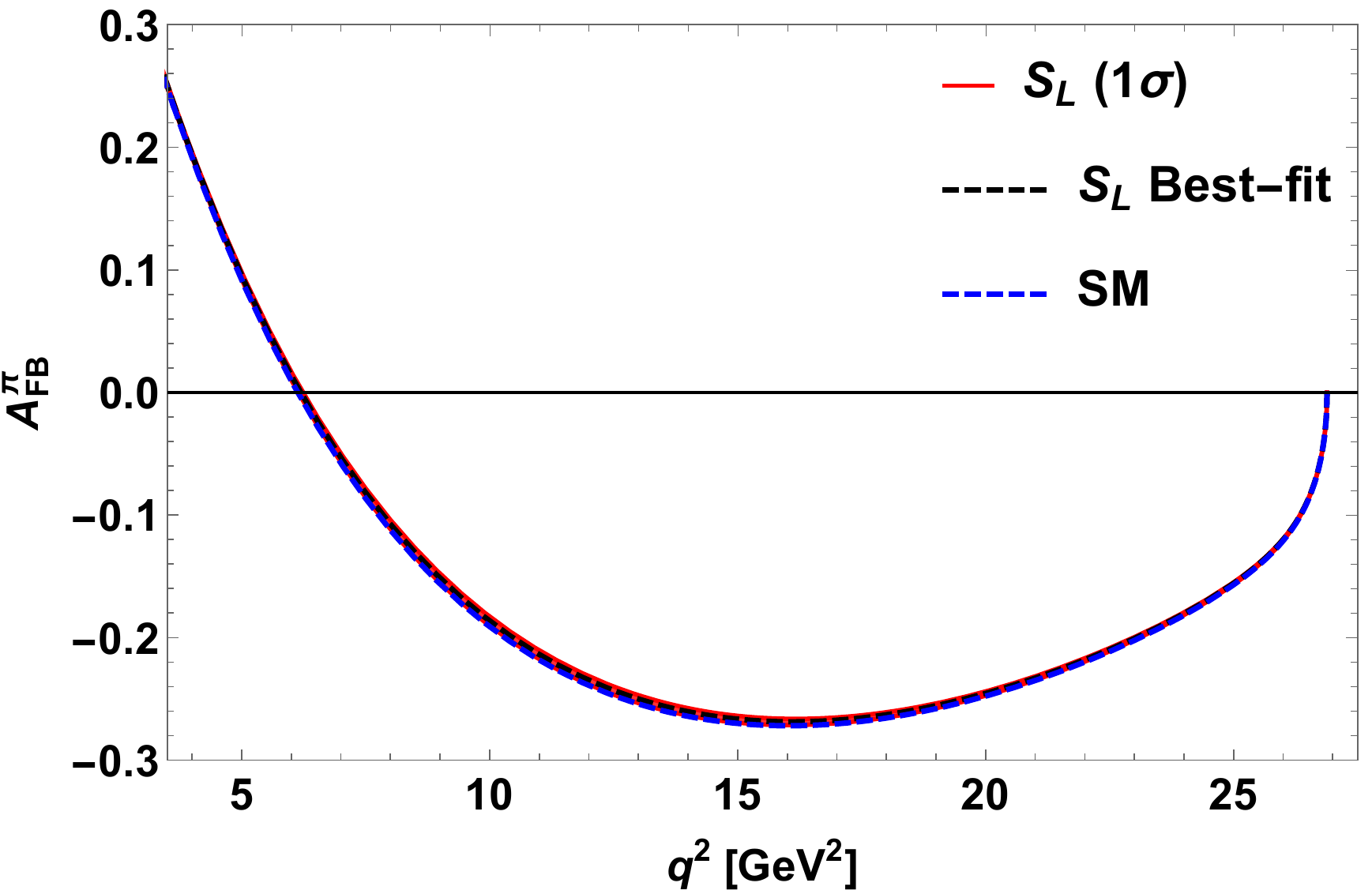}\\
\caption{The $q^2$ variation of  lepton spin asymmetry and forward-backward asymmetry of $ \bar B_d^{*} \to D^+ \tau^- \bar{\nu_\tau}$ (left panel) and  $\bar{B}_d^* \to \pi^+ \tau   \bar{\nu}_\tau$ (right panel) in presence of $S_L$ coefficient only. The black dashed lines and the red bands are obtained by using the best-fit values and corresponding $1\sigma$ range of $S_L$ coefficient. }
\label{variation-SL}
\end{figure}

In this subsection, we consider the contribution of  $S_L$ new coefficient by  assuming that all other new Wilson coefficients have vanishing values.  As seen from Figs. \ref{Fig:bclnu-cntr} and \ref{Fig:bulnu-cntr}\,, the $S_L$ parameters are severely constrained by the current data. Within the allowed parameter space for $S_L$ coefficient presented in Table \ref{Tab:con}\,, we show  the $q^2$ variation of  lepton-spin asymmetry (top) and forward-backward asymmetry (bottom) of $B^* \to D^+ \tau \bar \nu$ ($B^* \to \pi^+ \tau \bar \nu$) process on the left panel (right panel) of Fig. \ref{variation-SL}\,. Here the plots obtained from the best-fit values ($1\sigma$ range) of $S_L$ coupling are represented by dashed black lines (red bands).   The numerical values of these observables are given in Table \ref{Tab:SLSR}\,. With the additional $S_L$ contribution, the deviation in the branching ratios and LNU observables  from their SM predictions are found to be minimal. 
Though the lepton spin asymmetry  and forward-backward asymmetry observables of $B^* \to D^+ \tau \bar \nu$ channel provide  slight deviation from their SM results, the deviation is negligible in the $B^* \to \pi^+ \tau \bar \nu$ modes.  The zero crossing point of the forward-backward asymmetry of $B^* \to D^+ \tau \bar \nu$ process shifted sightly towards the low $q^2$ region. The $A_{\rm FB}^{P}$ vanishing  values of $q^2$ predicted from the best-fit values and $1\sigma$ range of new $S_L$ coefficient are presented in Table \ref{Tab:zero-cross}\,. 

\begin{table}[htb]
\begin{center}
\caption{Predicted numerical values of differential decay rate,  LNU observables, lepton spin asymmetry and forward-backward asymmetry of $ \bar B_{d(s)}^{*} \to D^+(D_s^+) \tau^- \bar{\nu}_\tau$ and  $\bar{B}_{d,(s)}^* \to \pi^+ (K^+) \tau \bar{\nu}_\tau$ decay processes in presence of $S_{L,R}$ coefficients. }\label{Tab:SLSR}
\begin{tabular}{| c |c |  c|}
\hline
Observables~  & Values with  $S_L$ & Values with $S_R$   \\
 \hline
 \hline
 $\rm{Br}(B^{*0} \to D^+ \tau^- \bar{\nu}_\tau)$  &
 $[2.193, 3.344] \times 10^{-8}$~&   $[2.180,3.251] \times 10^{-8}$\\
 
  $R^*_D$  & $[0.296,  0.298] $ & $[0.290, 0.294]$ \\
  
  $A_\lambda^D$  & $[0.581, 0.594]$ & $[0.604, 0.626]$\\
$A_{\rm FB}^D$  & $[-0.066, -0.058]$ & $[-0.126, -0.096]$\\  
 \hline
$\rm{Br}(B_s^{*} \to D_s^+ \tau^- \bar{\nu}_\tau)$  &
 $[3.993,6.089] \times 10^{-8}$&   $[3.968, 5.916] \times 10^{-8}$\\
  $R^*_{D_s}$  & $[0.293,0.296] $ & $[0.287, 0.292]$ \\
  $A_\lambda^{D_s}$  & $[0.578, 0.591]$ & $[0.601, 0.624]$\\
$A_{\rm FB}^{D_s}$  & $[-0.065, -0.056]$ & $[-0.126,  -0.095]$\\  
 \hline 
  $\rm{Br}(B^{*0} \to \pi^+ \tau^- \bar{\nu}_\tau)$  &~
 $[0.736, 1.285] \times 10^{-9}$~& ~  $[0.719, 1.250] \times 10^{-9}$~\\
 $R^*_\pi$  &  $[0.678, 0.680]$ & $[0.662, 0.663]$ \\
 $A_\lambda^{\pi}$  & $[0.774, 0.780]$ & $[0.822, 0.823]$\\
$A_{\rm FB}^\pi$  & $[-0.208, -0.204]$ & $[-0.254, -0.251]$\\  
 \hline 
  $\rm{Br}(B_s^{*} \to K^+ \tau^- \bar{\nu}_\tau)$   &
 $[0.755, 1.320] \times 10^{-9}$&   $[0.732, 1.273] \times 10^{-9}$\\
 $R^*_K$  & $[0.640, 0.642]$  & $[0.619, 0.620]$ \\
 $A_\lambda^{K}$  & $[0.738, 0.746]$&$[ 0.800, 0.802]$\\
$A_{\rm FB}^K$  & $[-0.207,  -0.202]$ & $[-0.260, -0.256]$\\ 
 \hline
\end{tabular}
\end{center}
\end{table}

\subsection{Effect of $S_R$ only}

\begin{figure}
\includegraphics[scale=0.4]{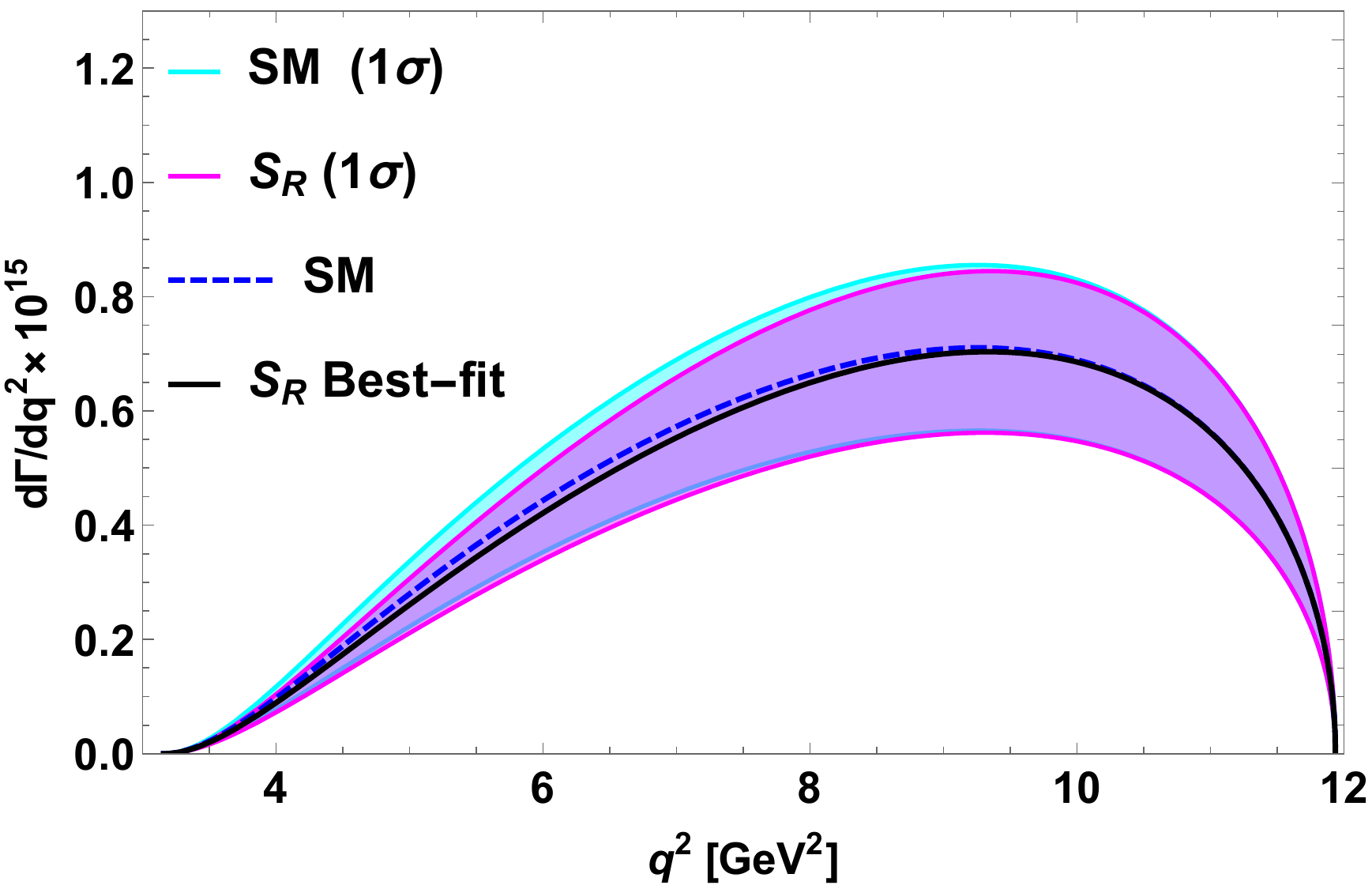}
\quad
\includegraphics[scale=0.4]{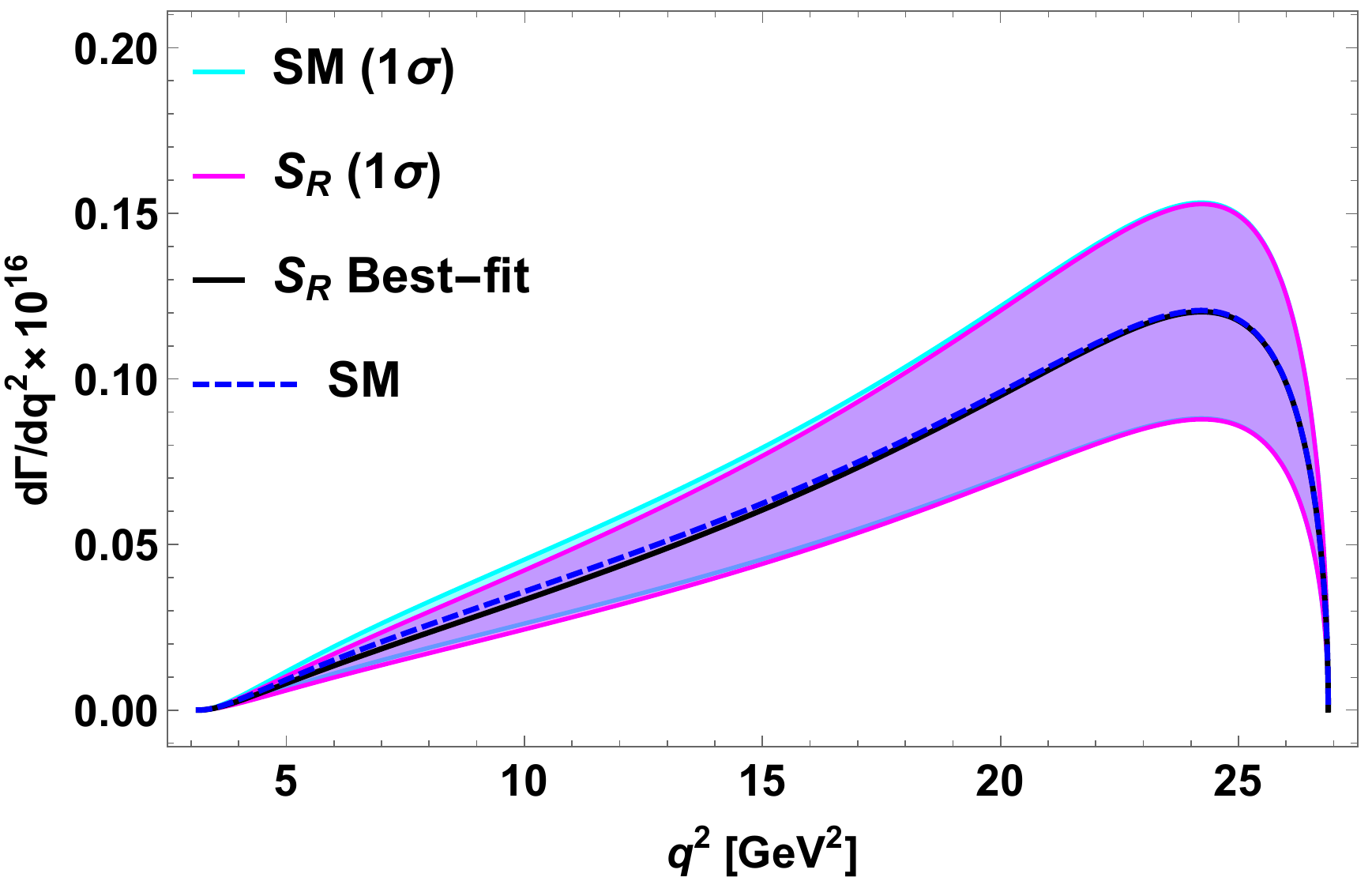}
\quad
\includegraphics[scale=0.4]{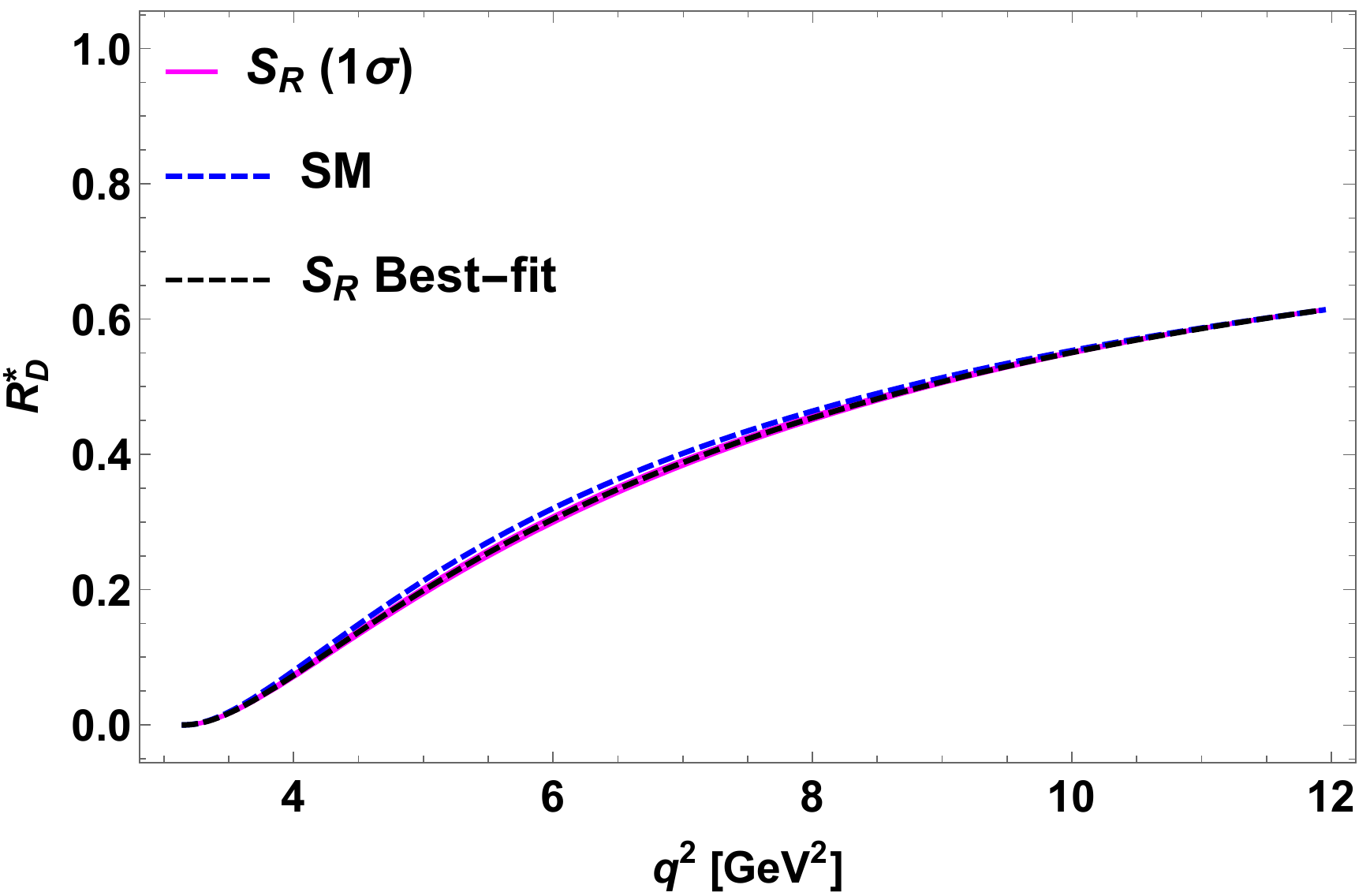}
\quad
\includegraphics[scale=0.4]{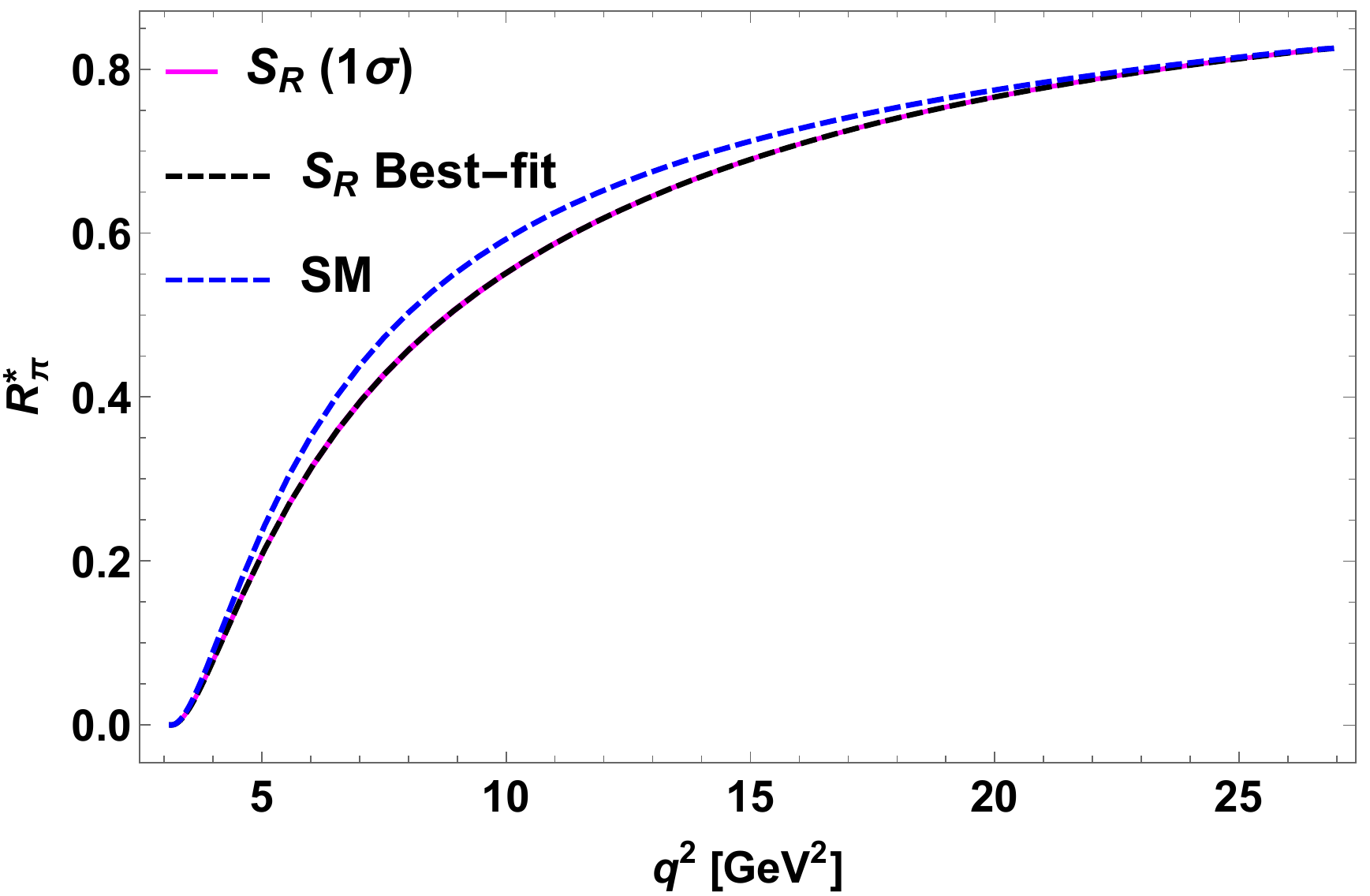}
\quad
\includegraphics[scale=0.4]{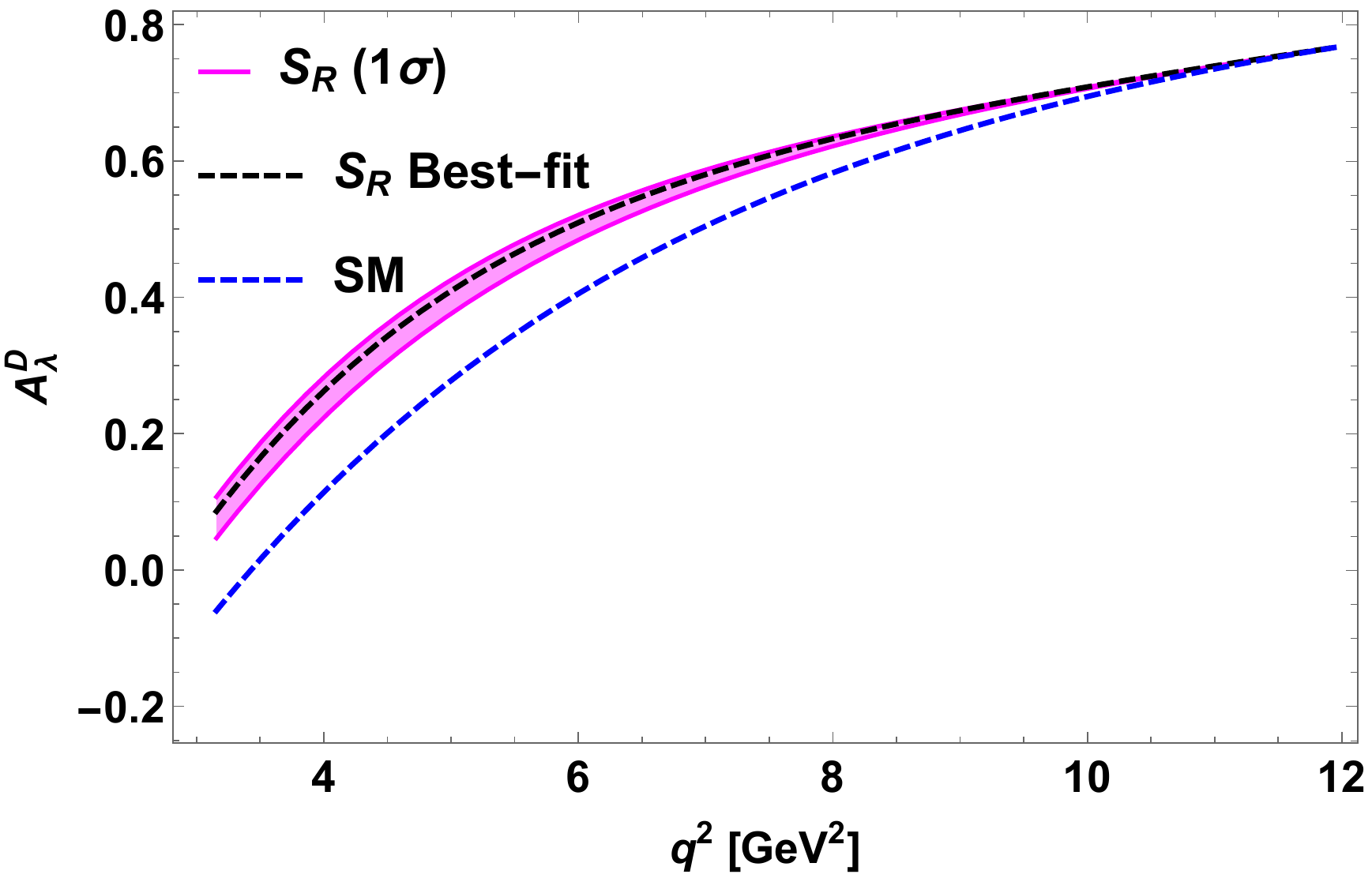}
\quad
\includegraphics[scale=0.4]{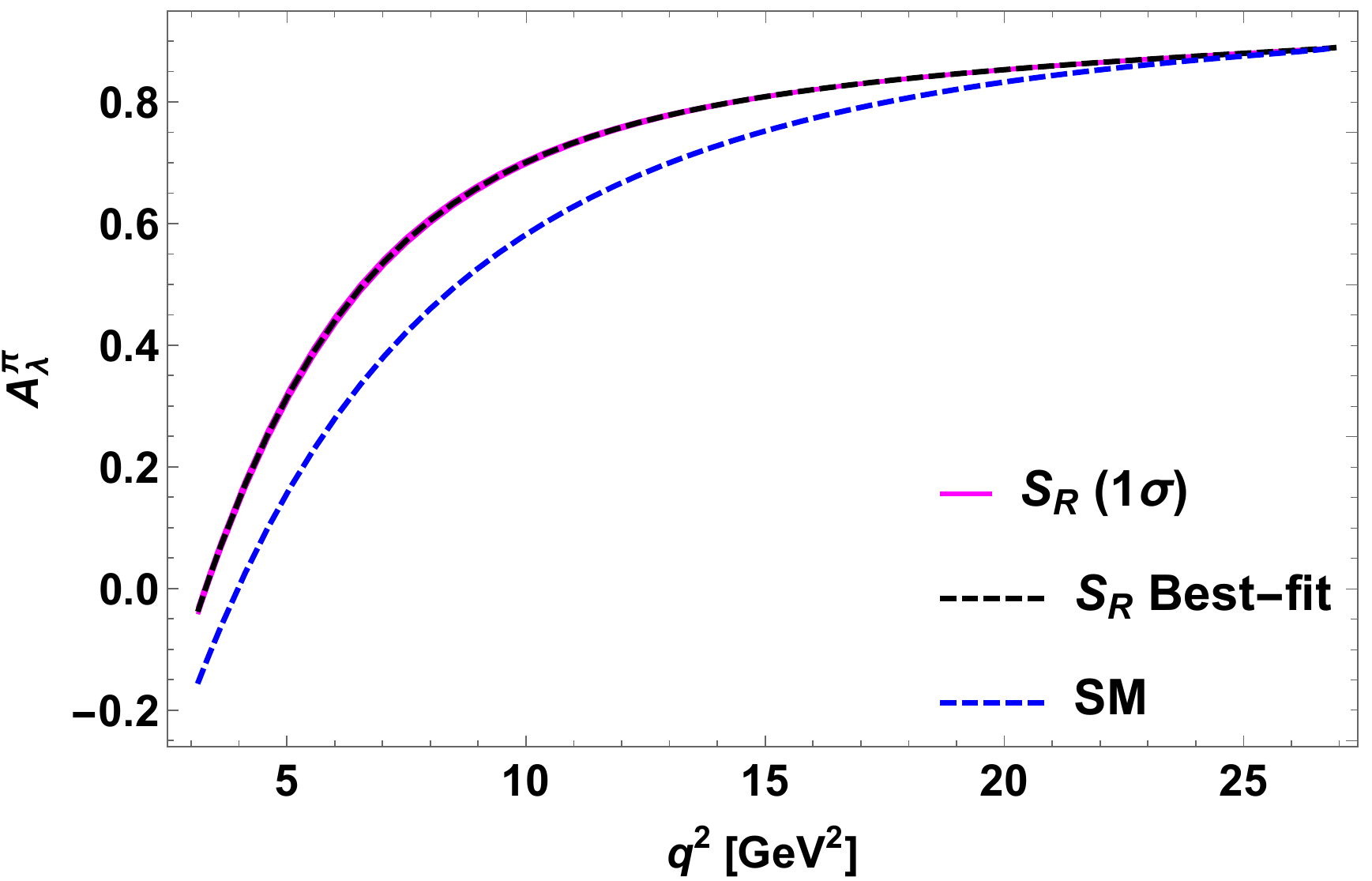}
\quad
\includegraphics[scale=0.4]{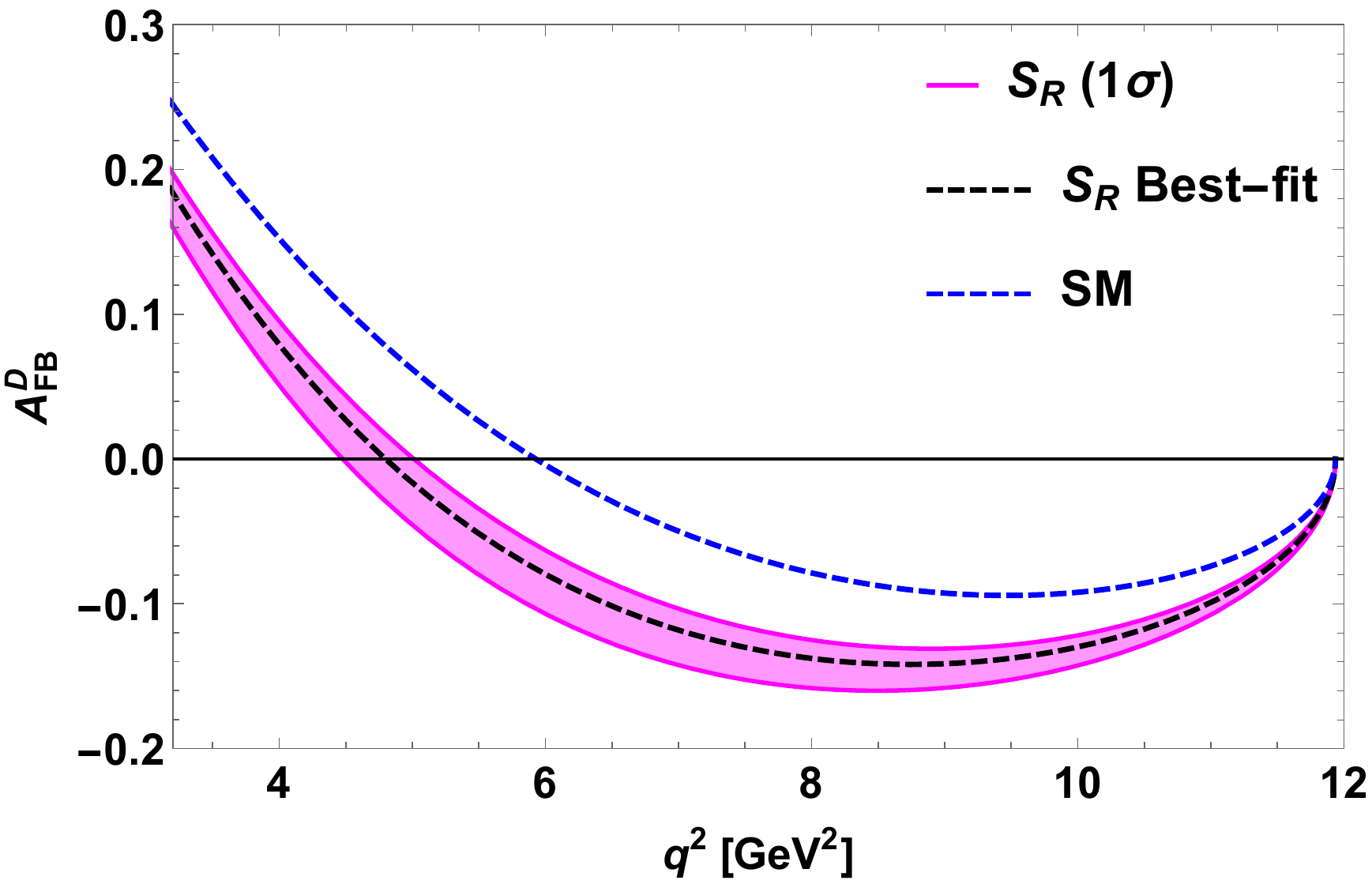}
\quad
\includegraphics[scale=0.4]{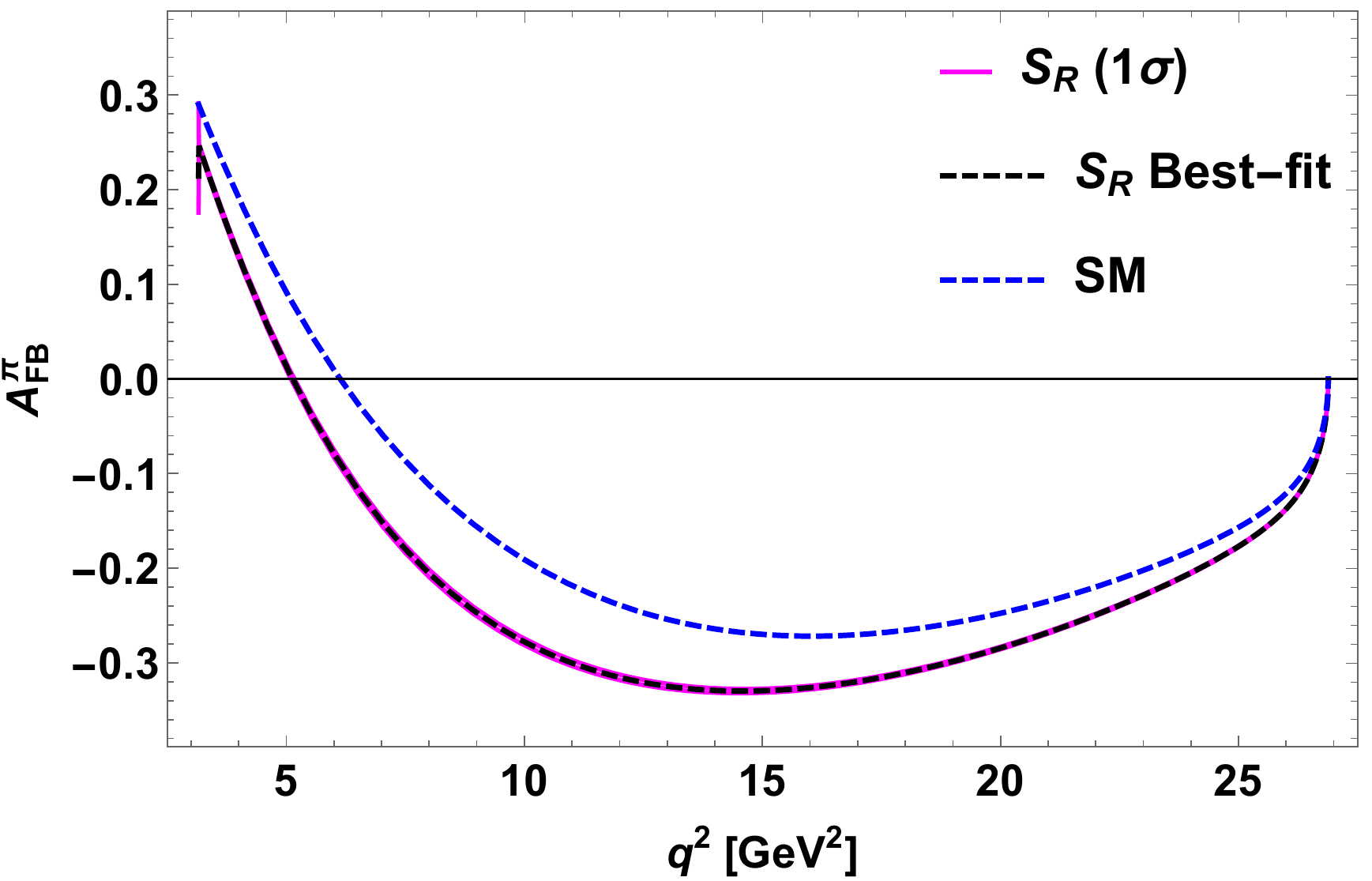}
\caption{The $q^2$ variation of differential decay rate, LNU observable, lepton spin asymmetry and forward-backward asymmetry of $ \bar B_d^{*} \to D^+ \tau^- \bar{\nu}_\tau$ (left panel) and  $\bar{B}_d^* \to \pi^+ \tau \bar{\nu}_\tau$ (right panel) in presence of $S_R$ coefficient only. The black dashed lines and the magenta bands are obtained by using the best-fit values and corresponding $1\sigma$ range of $S_R$ coefficient. }.
\label{variation-SR}
\end{figure}

Here we investigate the observables of $B^* \to (D^+, \pi^+) \tau \bar \nu$ decay modes by considering the presence of only additional $S_R$  coefficient. Using the available experimental data on $b \to (u,c) \tau \bar \nu$ transitions, we fit the corresponding $S_R$ coefficients, which is already discussed in section II.  In the left panel of Fig. \ref{variation-SR}\,, we present the $q^2$ variation of decay rate (top), $R^*_D$ (second from top), lepton spin asymmetry (third from top) and forward-backward asymmetry (bottom) of $B^* \to D^+ \tau \bar \nu$ and the corresponding plots for $B^* \to  \pi^+ \tau \bar \nu$ are shown in the right panel. Here the black dashed lines (magenta bands) are   obtained from the best-fit values ($1\sigma$ range) of $S_R$ coupling and other input parameters.   In this case also,  the deviation in the lepton spin asymmetry and forward-backward asymmetry observables are comparatively large, whereas the deviations in the branching ratios and LNU observables are nominal. The numerical values are presented in Table \ref{Tab:SLSR}\,. From Fig. \ref{variation-SR}\,, one can notice that the zero crossing point of the forward-backward asymmetry deviates significantly towards left (low $q^2$ region) and the corresponding $q^2$ values of the crossings are shown in Table \ref{Tab:zero-cross}\,.

\begin{table}[htb]
\begin{center}
\caption{The $q^2$ values (in GeV$^2$) of the zero crossing of forward-backward asymmetries of $B_{d,s}^* \to P \tau \bar \nu_\tau$ decay modes in the SM and in the presence of individual $V_{R}$, $S_{L,R}$ coefficients. The presence of additional $V_L$ coefficient don't change the $q^2$ crossing values of the $A_{\rm FB}^P$. }\label{Tab:zero-cross}
\begin{tabular}{| c |c |  c| c| c|}
\hline
Model~  &~ $B_d^* \to D \tau \bar \nu_\tau$ ~&~ $B_d^* \to \pi \tau \bar \nu_\tau$~ &~ $B_s^* \to D_s \tau \bar \nu_\tau$ ~&~$B_s^* \to K \tau \bar \nu_\tau$ ~\\
 \hline
 \hline

~  SM ~& ~$5.93$~ &~ $6.13$~&~5.96 ~&~6.26 ~\\
\hline
~ $V_R$ Only (Best-fit) ~& ~$6.88$~  &~ $6.88$~&6.92 & 7.03\\
~~~~~~~~~~~($1\sigma$) ~& ~$[6.56, 7.25]$ ~ & ~$[6.28,7.54]$~& $[6.59,7.28]$& $[6.42,7.70]$\\
\hline
~ $S_L$ Only (Best-fit) ~&~ $5.75$ ~ &~ $6.19$~& 5.78&6.33 \\
~~~~~~~~~~~($1\sigma$) ~&~ $[5.66,5.85]$~  & ~$[6.14,6.23]$~&~$[5.69,5.88]$ &$[6.28,6.37]$ \\
\hline
~ $S_R$ Only (Best-fit) ~&~ $4.80$  ~&~ $5.13$~& 4.82&5.22 \\
~~~~~~~~~~~($1\sigma$) ~&~ $[4.48,5.01]$  ~&~ $[5.09,5.17]$~&$[4.49,5.03] $&
$[5.18,5.26]$ \\
\hline
 
 \hline
\end{tabular}
\end{center}
\end{table}

\section{Summary and Conclusion}

The rare decay modes of $B$ mesons have been extensively studied  both theoretically and experimentally in order to critically test the standard model prediction and to look for new physics beyond it. In this regard, the  rare decay channels of the corresponding vector mesons i.e., the  $B^*$ decay modes are essential as they can provide  complementary ways to go beyond the standard model. However,  the weak decay channels of  $B^*$ vector mesons are not much explored experimentally as they decay dominantly through electromagnetic process $B^* \to B \gamma$. Recently, with the advent of high luminosity LHCb experiment the sensitivity for the branching fractions of various rare decay modes is expected to reach the level  $\sim \mathcal{O}(10^{-9})$. Thus, the LHCb would be an ideal platform to explore the rare decay modes of $B^*$ mesons. 

In view of the recently observed anomalies $R_{D^{(*)}}, R_{J/\psi}, R_\pi^l$ involving  the charged current $b\to (c,u)l \nu$ transitions, we  have performed a model independent analysis of the semileptonic decay process of $B^*$ vector meson decaying to a pseudoscalar meson $P$, where $P=D, D_s,\pi, K$, along with a charged lepton and corresponding antineutrino. We  considered the generalized effective Lagrangian in  the presence of  vector and scalar   type new physics operators.  
Considering only one new coefficient to be present  at a time, and  assuming  the new couplings as complex, we  constrained the  new parameters associated  with $b \to c \tau \bar \nu_\tau$ processes by  performing $\chi^2$ fit from $R_{D^{(*)}}$, $R_{J/\psi}$ parameters and the upper limit on $B_c^+ \to \tau^+ \nu_\tau$ branching fraction. The new couplings of $b \to u \tau \bar \nu_\tau$ processes are constrained by using experimental data on the branching ratios of $B_u \to \tau \nu_\tau$ and $B \to \pi \tau \nu_\tau$ and $R_\pi^l$ parameter. Using the best-fit values and the corresponding $1\sigma$ ranges of new  individual complex Wilson coefficients, we  computed the branching ratios, forward-backward asymmetry, lepton spin asymmetry and lepton non-universality observables of $ B_{d(s)}^{*} \to D^+(D_s^+) \tau^- \bar{\nu}_\tau$ and $\bar{B}_{d(s)}^* \to \pi^+(K^+) \tau\bar{\nu}_\tau$ decay processes.   We have also shown the values of $q^2$ at which the forward-backward asymmetry  vanishes.  The branching fractions and LNU observables of these decay modes in the presence of additional $V_L$ coefficient have  significant deviations from their corresponding standard model predictions, whereas no deviations have  been found in the lepton spin asymmetry and forward-backward asymmetry observables. Due to the additional contributions from $V_R$ new coefficient, profound deviations have observed in the decay rates, lepton nonuniversality observable and the forward-backward asymmetry of both $\bar{B}^* \to (D, \pi) \tau \bar{\nu}_\tau$ processes. Due to the presence of $V_R$ coupling, the zero crossing of forward-backward asymmetry has shifted   towards high $q^2$ region for all decay modes.  In the presence of $S_L$ coefficient,  none of the observables are affected and there is practically no deviation from SM results.   Only the lepton-spin asymmetry and forward-backward asymmetry observables of $\bar{B}^* \to D \tau \bar{\nu}_\tau$ show slight deviation due to additional $S_L$ coupling. On the other hand, in the presence of $S_R$ coupling, the lepton spin asymmetry and the forward-backward asymmetry show reasonable deviations from their SM predictions and the decay rate, lepton nonuniversality  observables remain unchanged. The  zero crossing of forward-backward asymmetry of all decay modes in the presence of $S_R$ coefficient is found to be shifted  towards low $q^2$ region. 
To conclude, we noticed significant deviations in some of the  observables  from their standard model predictions  in presence of new couplings.  The observation of these decay modes of vector $B^*$ mesons at LHC experiment will definitely shed light on the nature of new physics.

\acknowledgements

RM  and AR would like to thank Science and Engineering Research Board (SERB), Government of India for financial support through grant No. EMR/2017/001448.

\bibliography{BL}

\end{document}